\documentclass[]{jfm}
\usepackage{graphicx}
\usepackage{newtxtext}
\usepackage{newtxmath}
\usepackage{natbib}
\usepackage{hyperref}
\hypersetup{
    colorlinks = true,
    urlcolor   = blue,
    citecolor  = black,
}

\newcommand{\RomanNumeralCaps}[1]
\linenumbers

\title{Direct numerical simulations of optimal thermal convection in rotating plane layer dynamos.}

\author{Souvik Naskar\aff{1}, 
  Anikesh Pal\aff{1}
  \corresp{\email{pala@iitk.ac.in}}}

\affiliation{\aff{1}Department of Mechanical Engineering, Indian Institute of Technology, Kanpur 208016, India}

\begin{document}
\maketitle

\begin{abstract}
The heat transfer behavior of convection-driven dynamos in a rotating plane layer between two parallel plates, heated from the bottom and cooled from the top, is investigated. At a fixed rotation rate (Ekman Number, $E=10^{-6}$) and fluid properties (thermal and magnetic Prandtl numbers, $Pr=Pr_m=1$), both dynamo convection (DC) and non-magnetic rotating convection (RC) simulations are performed to demarcate the effect of magnetic field on heat transport at different thermal forcing (Rayleigh Number, $Ra=3.83\times10^{9}-3.83\times10^{10}$). In this range, our turbulence resolving simulations demonstrate the existence of an optimum thermal forcing, at which heat transfer between the plates in DC exhibits maximum enhancement, as compared to the heat transport in the RC simulations. Unlike any global force balance reported in the literature, the present simulations reveal an increase in the Lorentz force in the \textit{thermal boundary layer}, due to stretching of magnetic field lines by the vortices near the walls with no-slip boundary condition. This increase in Lorentz force mitigates turbulence suppression owing to the Coriolis force, resulting in enhanced turbulence and heat transfer.
\end{abstract}

\begin{keywords}
\end{keywords}


\section{Introduction}\label{sec:intro}

Hydromagnetic dynamo action is the commonly accepted source of sustained magnetic fields in stars and planets. In this mechanism, a convective system with electrically conducting fluid can amplify a small magnetic field and continuously convert the kinetic energy of the fluid to magnetic energy, to sustain the field against Joule dissipation. This self-exciting dynamo action can be facilitated due to the presence of rotation leading to large-scale organized fields \citep{moffatt_2019}. The most familiar example of such a rotating convection-driven dynamo is the Earth's outer core, where the dynamo mechanism originating from the turbulent motions of electrically conducting liquid iron in Earth's outer core forms the sustained dipole dominant geomagnetic field \citep{roberts_2013}. Thermal convection in rotating dynamos, both in plane layer \citep{stellmach_2004,tilgner_2012} and spherical \citep{glatzmaiers_1995,christensen_2015} geometries, have been extensively investigated to understand their force balance, nature of MHD turbulence, and magnetic field behavior. Another aspect of fundamental importance is the heat transfer to diagnose the global transport properties of such systems. Convective heat transfer through the atmosphere decides the weather, climate, planetary and stellar energetics, their evolution \citep{schmitz_2009}, and generation of magnetic fields from planetary cores \citep{Nimmo_2015}. However, heat transport in self-excited rotating dynamos is yet to be explored.\\

Dynamics of flow and heat transfer in thermal convection with rotation and magnetic field depends on four governing non-dimensional numbers: (I)  the Rayleigh Number ($Ra$) representing the thermal forcing, (II) the Ekman Number($E$) signifying the ratio of viscous force to Coriolis force, (III) the thermal Prandtl number ($Pr$) and (IV) the magnetic Prandtl number ($Pr_m$). $Pr$ and $Pr_m$ are fluid properties as defined in equation \ref{eqn:nd_parameters}. These four numbers decide the global characteristics of the system, indicated by (a) the magnetic Reynolds Number ($Re_m = u_\textit{r.m.s}d/\lambda$, where $u_\textit{r.m.s}$ and $d$ are the characteristics velocity and length scales, whereas $\lambda$ is the magnetic diffusivity), characterizing the dominance of electromagnetic induction over ohmic diffusion of the magnetic field, (b) the Nusselt Number ($Nu$) representing non-dimensional heat transfer(see equation \ref{eqn:nu}) and, (c) the Elsasser Number ($\Lambda=\sigma B_\textit{r.m.s}^2/\rho\Omega$, where $\sigma$ is the electrical conductivity, $B_\textit{r.m.s}$ is the characteristics magnetic field strength, $\rho$ is the density, and $\Omega$ is the rotation rate), depicting the ratio of the Lorentz and the Coriolis forces.\\  

Based on the behavior of magnetic field with varying $Re_m$, thermal convection with rotation and magnetic field can be broadly classified into two categories: (I) rotating magnetoconvection (RMC) at low magnetic Reynolds number ($Re_m\xrightarrow{}0$) and (II) dynamo convection(DC) for moderate to high values $Re_m\ge{}O(10)$. For RMC, the system can not self-induce or sustain a magnetic field, and therefore the magnetic field has to be imposed externally. As the self-induced magnetic field remains small with respect to the imposed field, linear analysis may be used to predict the critical value of Rayleigh number ($Ra_c$) at the onset of convection \citep{chandrasekhar_1961}. Moreover, liquid metals (e.g. Mercury, Sodium, Gallium), having large magnetic diffusivities $(\lambda \xrightarrow{}O(10^6-10^7))$ are used to conduct RMC experiments at low $Re_m$. Strong rotation in any convective system restricts convection, attributed to the Taylor-Proudman constraint imposed by the Coriolis force, which inhibits changes along the axis of rotation. Linear theory of RMC \citep{chandrasekhar_1961} suggests that magnetic field can relax Taylor-Proudman constraint to enhance convection, known as the "magnetorelaxation" process. These theoretical predictions have been verified by experiments \citep{nakagawa_1957,nakagawa_1959,aurnou_2001}. \citet{eltayeb_1972} performed an extensive linear analysis of RMC in the asymptotic limits of large rotation rate ($E\xrightarrow{}0$) with large imposed field strength for various boundary conditions, orientations of the external magnetic field, and rotation. Later, the linear theory was extended to allow for finite-amplitude modes considering the dominant modes contribute to most of the heat transport \citep{stevenson_1979}. In RMC, the action of Lorentz force on the flow can lead to enhanced heat transfer \citep{king_2015} as compared to non-magnetic rotating convection(RC). The highest enhancement occurs when the Lorentz force due to the magnetic field and the Coriolis force due to rotation balance each other, as found in the experiment of \citet{king_2015}. This global balance, known as the "magnetostrophic" state, is characterized by Elsasser number of order unity, $\Lambda \approx O(1)$.\\

Challenges in producing dynamo action in laboratory experiments \citep{plumley_2019} compel the studies on turbulent dynamo convection (DC) to depend primarily on numerical simulations. Self-exciting dynamo simulations at moderate values of magnetic Reynolds number $(Re_m = 243-346)$ in a spherical shell model performed by \citet{Yadav_2016}, revealed a fundamentally different force balance compared to the experimental results of \citet{king_2015}. In such "geodynamo" simulations that attempt to model the Earth's outer core, the primary force balance is "geostrophic" between Coriolis and pressure forces, whereas the Magnetic, the Archemedian (buoyancy), and the Coriolis forces(MAC) constitute a lower order of magnitude (quasi-geostrophic, QG) balance, abbreviated as the QG-MAC balance \citep{aubert_2019,schwaiger_2019}. Interestingly, \citet{Yadav_2016} also reported a similar heat transfer enhancement as found in the RMC experiments of \citet{king_2015}, irrespective of very different force balance and parameter regime. \\ 

The present study explores the heat transfer behavior in DC with a plane layer geometry using direct numerical simulations (DNS). This is the simplest possible model of DC in a rotating layer of electrically conducting fluid between two parallel plates, heated from the bottom and cooled from the top (see figure \ref{fig:domain}a). Using this model, \citet{jones_2000} investigated the effect of $Ra$, $E$, and $Pr_m$ on the growth and saturation of the magnetic field and the global force balance while neglecting inertial effects at large Prandtl numbers ($Pr\xrightarrow{}\infty$). The dynamo was found to saturate at $\Lambda\approx O(1)$. At moderate values of magnetic Reynolds number, $Re_m\approx O(10^2)$ a simple model of dynamo convection was proposed. \citet{rotvig_2002} studied DC in the range $E=10^{-4}-10^{-5}$ and reported a MAC balance in the bulk with $\Lambda \approx O(10)$. Rapidly rotating ($E=2\times10^{-4}-10^{-6}$), weakly non-linear DC was explored by \citet{stellmach_2004}. They reported time-dependent flow with cyclic variation between small and large-scale magnetic fields. Interestingly, dynamo action was found to exist even below the critical Rayleigh number for the onset of non-magnetic convection($Ra_c$), indicating subcritical dynamo action. \citet{cattaneo_2006} analyzed the $\alpha$-effect (a mechanism of large-scale field generation from small-scales due to the helical nature of the flow in DC \citep{moffatt_2019}) and reported small-scale dynamos with a diffusively controlled $\alpha$-effect, without any indication of large-scale field generation. A transition between small-scale and large-scale field generation was found by \citet{tilgner_2012,tilgner_2014} with separate scaling laws for the magnetic energy. The transition occurs at $Re_mE^{1/3}=13.5$, below which the magnetic field generation is dependent on flow helicity. Equipartition of energies is observed above the transition point where the field stretching process at high $Re_m$ generates small-scale fields independent of flow helicity.  \citet{guervilly_2015} demonstrated the existence of large-scale vortices in DC that can lead to large-scale field generation, which is of great interest for astrophysical dynamos. These vortices reduced the heat transfer efficiency. However, for high $Re_m$, generation of small-scale magnetic field suppresses the formation of such vortices \citep{guervilly_2017}. \citet{calkins_2015} investigated asymptotically reduced DC models with leading order QG balance. Four distinct QG dynamo regimes have been identified with varying dynamics and magnetic to kinetic energy density ratios \citep{calkins_2018}. Extension to non-Boussinesq convection was investigated by \citet{kapyla_2009} and \citet{favier_2013}. \\ 

The heat transport behavior in plane layer DC remains almost unexplored. In this geometry, two distinct regions of convection can be identified: the boundary layer region near the plates with high velocity and temperature gradients (Ekman and thermal boundary layers) and a well-mixed "bulk" regime in the interior where gradients are much smaller. Heat transfer scaling of dynamos without rotation has been found to be identical to non-rotating, non-magnetic Rayleigh B\'enard convection \citep{yan_2021}, where heat transfer is known to be constrained by the boundary layer dynamics \citep{plumley_2019}. Conversely, heat transfer in non-magnetic rotating convection is constrained by the bulk, though it can be significantly altered by the boundary layer dynamics \citep{stellmach_2014}. Even without considering the viscous boundary layer, \citet{stellmach_2004} found enhanced heat transfer in weakly non-linear convection in a rapidly rotating dynamo. They attributed this increased heat transport to the increased length scale of convection due to the presence of a strong magnetic field. However, the variation of heat transfer with thermal forcing in a rotating dynamo remains unknown. Also, to the best of our knowledge, no studies report the role of the boundary layer on the heat transfer behavior in rotating DC. We perform a systematic study,  unprecedented to the previous investigations, in a plane layer geometry to investigate the boundary layer dynamics of dynamos and its implications on the heat transfer behavior.\\  

\section{Method}\label{sec:method}
\subsection{Governing equations and numerical details}\label{sec:govequ}

\begin{figure}
\centering
(a)\includegraphics[width=0.63\textwidth,trim={0 2cm 0 1cm},clip]{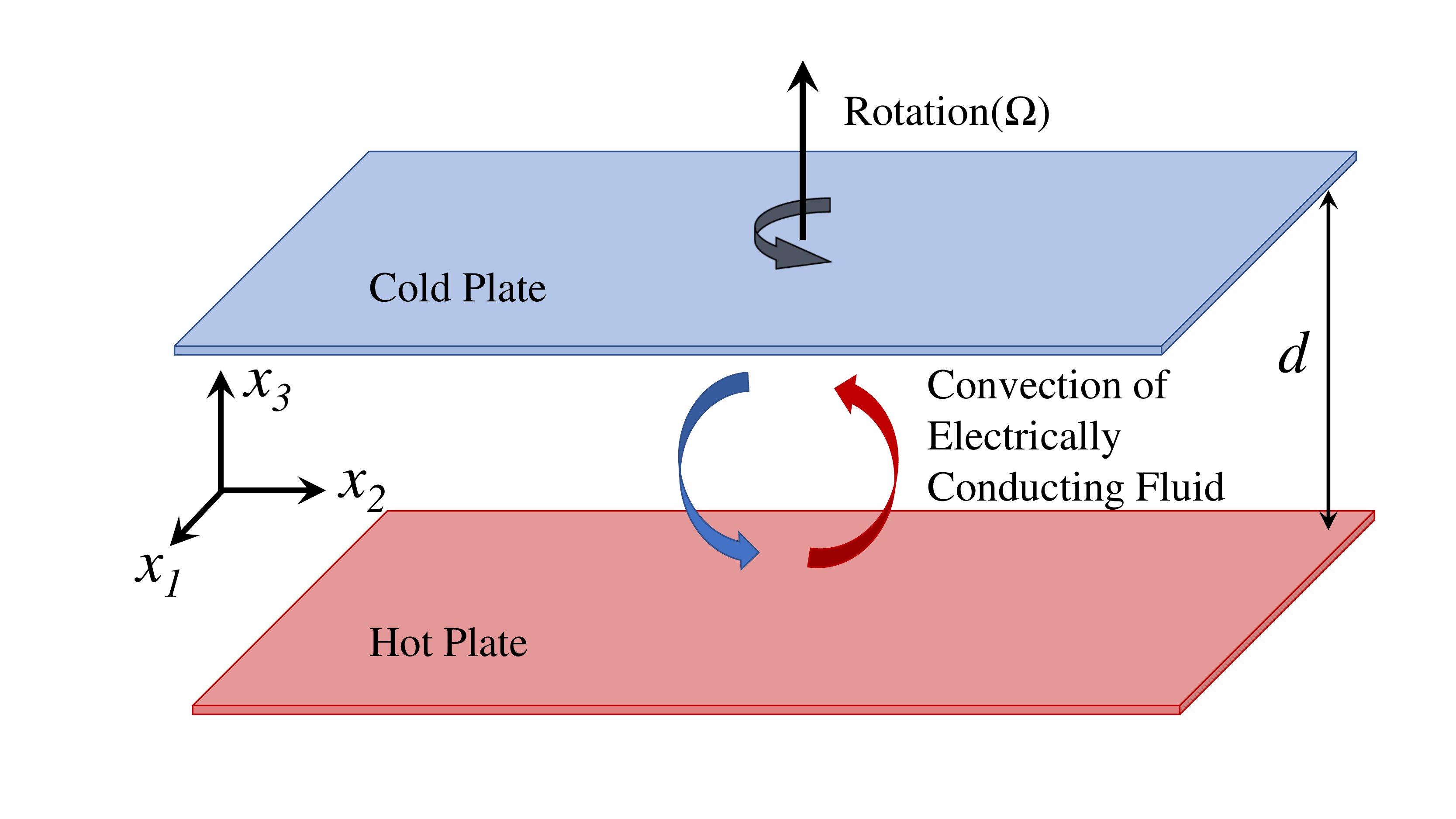}
(b)\includegraphics[width=0.3\textwidth,trim={0cm 4cm 0 4cm},clip]{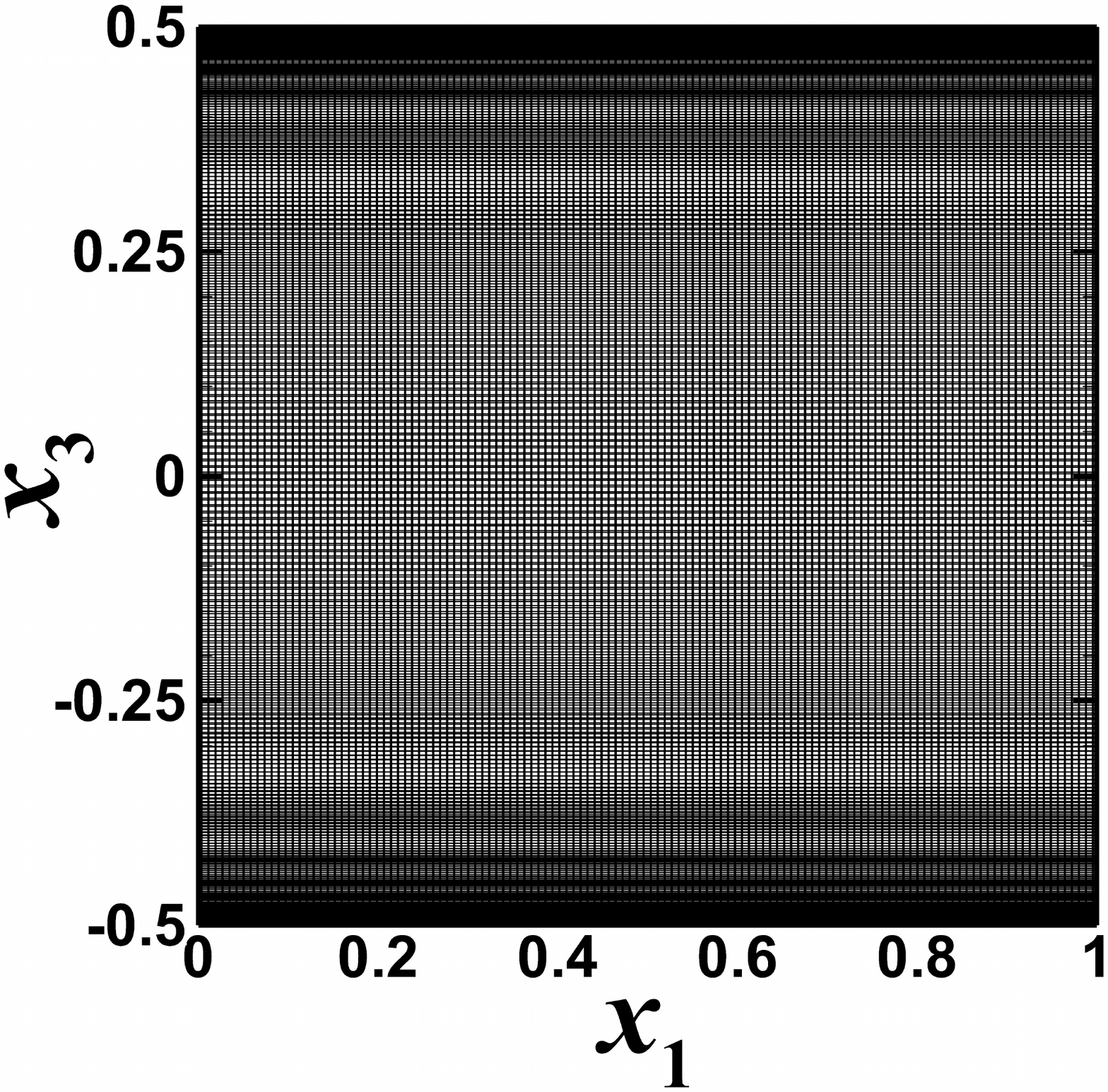}
\caption{(a) Plane layer dynamo model (b) Grid clustering near the walls to resolve boundary layers}
\label{fig:domain}
\end{figure}

We consider dynamo convection in a three-dimensional Cartesian layer of incompressible, electrically conducting, Boussinesq fluid. The layer of depth $d$ is kept between two parallel plates with temperature difference $\Delta T$ where the lower plate is hotter (see figure \ref{fig:domain}a). The system is rotating about the vertical with an angular velocity $\Omega$. The fluid properties are density($\rho$), kinematic viscosity($\nu$), thermal diffusivity($\kappa$), magnetic permeability($\mu$), electrical conductivity($\sigma$) and the magnetic diffusivity($\lambda$). The layer depth $d$ is the length scale, free-fall velocity, $u_f=\sqrt{g\alpha\Delta T d}$, is the velocity scale, $\sqrt{\rho\mu}u_f$ is the magnetic field scale and the temperature drop across the layer, $\Delta T$ is the temperature scale assumed for non-dimensionalizing the governing equations \citep{iyer_2020}. The non-dimensional equations for the velocity field $\boldsymbol{u}$, temperature field $\theta$, and magnetic field $\boldsymbol{B}$, take the following form:

\begin{equation}\label{eqn:solenoidal_nd}
    \frac{\partial u_j}{\partial x_j}=
    \frac{\partial B_j}{\partial x_j}=0,
\end{equation}
\begin{equation}\label{eqn:momentum_nd}
\begin{split}
    \frac{\partial u_i}{\partial t}+
    u_j\frac{\partial u_i}{\partial x_j}=
    -\frac{\partial p}{\partial x_i}+
    \frac{2}{E}\sqrt{\frac{Pr}{Ra}}\epsilon_{ij3} u_j\hat{e_3}+
    B_j\frac{\partial B_i}{\partial x_j}+
    \theta\delta_{i3}+
    \sqrt{\frac{Pr}{Ra}}\frac{\partial^{2} u_{i}}{\partial x_j\partial x_j},
\end{split}
\end{equation}
\begin{equation}\label{eqn:energy_nd}
    \frac{\partial \theta}{\partial t}+
    u_j\frac{\partial \theta}{\partial x_j}=
    \frac{1}{\sqrt{RaPr}}\frac{\partial^2\theta}{\partial x_j\partial x_j},
\end{equation}
\begin{equation}\label{eqn:maxwell_nd}
    \frac{\partial B_i}{\partial t}+
    u_j\frac{\partial B_i}{\partial x_j}=
    B_j\frac{\partial u_i}{\partial x_j}+
    \sqrt{\frac{Pr}{Ra}}\frac{1}{Pr_m}\frac{\partial^{2} B_{i}}{\partial x_j\partial x_j}.
\end{equation}

The non-dimensional parameters appearing in equations [\ref{eqn:momentum_nd}-\ref{eqn:maxwell_nd}] are defined below:
\begin{equation}\label{eqn:nd_parameters}
    Ra=\frac{g\alpha\Delta T d^3}{\kappa\nu},\quad E=\frac{\nu}{\Omega d^2},\quad Pr=\frac{\nu}{\kappa},\quad Pr_m=\frac{\nu}{\lambda}. 
\end{equation}

In the horizontal directions ($x_{1},x_{2}$) periodic boundary conditions are applied. The velocity boundary conditions are no-slip and impenetrable. As we aim to study the effect of boundary layer dynamics on dynamo convection, the no-slip boundary condition is the natural choice \citep{stellmach_2014,jones_2000}. Although corresponding results for free-slip boundary condition are also presented, we discuss the results for no-slip boundary condition unless stated otherwise. Therefore,

\begin{equation}\label{eqn:ubc}
\begin{split}
    u_{1}=u_{2}=u_{3}=0 \; \textrm{ at } \; x_{3}=\pm1/2
\end{split}
\end{equation} 

An unstable temperature gradient is maintained by imposing isothermal boundary conditions.

\begin{equation}\label{eqn:tbc}
\begin{split}
    \quad \theta=1 \; \textrm{ at } \; x_{3}=-1/2, \quad \theta=0 \; \textrm{ at } \; x_{3}=1/2
\end{split}
\end{equation}

For the magnetic field, either perfectly conducting or perfectly insulating boundary conditions can be implemented. The insulating boundary condition is commonly applied in spherical shell geometries for geodynamo simulations \citep{Yadav_2016,christensen_2015}. However, for plane layer geometries, perfectly conducting boundary condition is commonly used \citep{tilgner_2012,guervilly_2015,hughes_2019} because it enables comparison with the theoretical studies \citep{stellmach_2004}. Laboratory setups for RMC \citep{king_2015} and dynamos \citep{monchaux_2007} also use conducting materials as containers of liquid metals that are commonly used to conduct experiments. Thus, perfectly conducting boundary condition are implemented in the present study such that,

\begin{equation}\label{eqn:mbc}
\begin{split}
    \frac{\partial B_{1}}{\partial x_{3}}=\frac{\partial  B_{2}}{\partial x_{3}}=0,\  B_{3}=0 \  \textrm{ at } \  x_{3}=\pm1/2.   
\end{split}
\end{equation} 

The governing equations are solved in a cubic domain with unit side length, using the finite difference method \citep{bakhuis_2018} in a staggered grid arrangement. The scalar quantities (pressure and temperature) are stored at the cell centers whereas, velocity and magnetic field components are stored at the cell faces. Second-order central difference is used for spatial discretization. The projection method is used to calculate the divergence-free velocity field where the pressure Poisson equation is solved using a parallel multigrid algorithm. Similarly, an elliptic divergence cleaning algorithm is employed to keep the magnetic field solenoidal \citep{Brackbill1980}. An explicit third-order Runge-Kutta method is used for time advancement except for the diffusion terms, which are solved implicitly using the Crank-Nicolson method \citep{Pham_2009,brucker_2010}. The solver has been validated extensively for numerous DNS of stratified turbulent flows \citep{Pal2013,Pal2015,pal_2020}.\\

Validation of our numerical solver is performed by replicating the results from the following literature: (I) critical Rayleigh number and wavenumber ($Ra_c$ and $a_c$) at the onset of non-rotating magnetoconvection with imposed field strength $Q=\sigma B^2d^2/\rho\nu=10^{-4}$ predicted from linear theory \citep{chandrasekhar_1961}, (II) RC simulations of \citet{stellmach_2004} at $Ra=1.4\times10^{7}$, $E=5\times10^{-5}$ and $Pr=1$, along with DC simulation results for $Pr_m=2.5$ with free-slip boundary conditions, (III) DNS results of quasi-static magnetoconvection \citep{yan_2019} at $Q=10^{-4}$ and $Ra=1.3\times10^{5}$. Our results are in good agreement with these earlier studies as demonstrated in table \ref{tab:validation}. \\

\begin{table}
  \begin{center}
\def~{\hphantom{0}}
  \begin{tabular}{ccccccccccc}
        &   $Ra_c$ &   $a_c$ &   $Nu_{q}$ &   $Re_{q}$ &   $Re_0$ &   $Nu_0$ &   $Re_m$ &   $Nu$ &   $\Lambda$ &   $ER$ \\[3pt]
       \cite{chandrasekhar_1961}   &  124509   &  8.66 &  - &  - &  - &  - &  - &  - &   & \\
       Validation I  &  124510   &  8.47 &   &   &   &   &   &   &   &  \\ 
       \cite{stellmach_2004}   &  -   &  - &  - &  - &  48.3 &  1.34 &  170.7 &  1.66 &  0.38 &  1.37\\
       Validation II &     &   &   &   &  48.6 &  1.36 &  168.5 &  1.68 &  0.36 & 1.36\\ 
       \cite{yan_2019}   &  -   &  - &  1.149 &  3.79 &  - &  - &  - &  -\\
       Validation III   &     &   &  1.149 &  3.72 &   &   &   &    &   & \\       
  \end{tabular}
  \caption{Results from the three test runs to reproduce results from literatures: (I) Linear magnetoconvection theory \citep{chandrasekhar_1961}, (II) RC and DC simulations of \citet{stellmach_2004} and (III) DNS results of quasi-static magnetoconvection \citep{yan_2019}. Subscript $"0"$ and $q$ are used to represent non-magnetic RC and quasi-static magnetoconvection results respectively. The magnetic to kinetic energy ratio is represented by $ER$.}
  \label{tab:validation}
  \end{center}
\end{table}

\subsection{Problem Setup}\label{sec:setup}

Dynamical balances and heat transport are studied at constant rotation rate and constant fluid properties with variation in thermal forcing. The thermal forcing here is represented by the convective supercriticality $\mathcal{R}=Ra/Ra_c$ where $Ra_c=19.15\times E^{-4/3}$ is the minimum required value of $Ra$ to start rotating convection \citep{chandrasekhar_1961} with no-slip boundary conditions (critical Rayleigh Number is higher for free-slip boundaries where $Ra_c=21.91\times E^{-4/3}$ for large rotation rates). We choose the values of $\mathcal{R}=2,2.5,3,4,5,10,20$, Ekman number $E=10^{-6}$ and the Prandtl Numbers $Pr=Pr_m=1$ for the present simulations. To elucidate the effect of the magnetic field, we perform two simulations at each value of $\mathcal{R}$: DC and RC. It should be noted here that the critical Rayleigh number for the onset of dynamo convection remains unknown \citep{plumley_2019}. Hence, we use the same value of $Ra_c$ for both DC and RC simulations following \citet{stellmach_2004}. In the horizontal directions ($x_1$ and $x_2$) $1024$ uniform grid points are used, whereas in the vertical direction ($x_3$) we have used $256$ non-uniform grid points. Capturing the boundary layer dynamics in these simulations is imperative for understanding the optimal heat transfer phenomenon, and requires a grid that can resolve the smallest dissipative scales. Grid clustering \citep{bakhuis_2018} near the walls (figure \ref{fig:domain}b) is performed to ensure a minimum of four grid points inside the Ekman layer with the spacing, $\Delta x_{3}=4.8\times 10^{-4}$ at the wall. This grid resolution is selected after subsequent grid refinements such that the kinetic and magnetic energies are sufficiently dissipated, and the turbulent kinetic energy budget closure is maintained. The simulation details and global diagnostic parameters are summarized in table \ref{tab:volavg}.\\

\begin{figure*}
\centering
(a)\includegraphics[width=\textwidth,trim={0cm 10.5cm 0 12.5cm},clip]{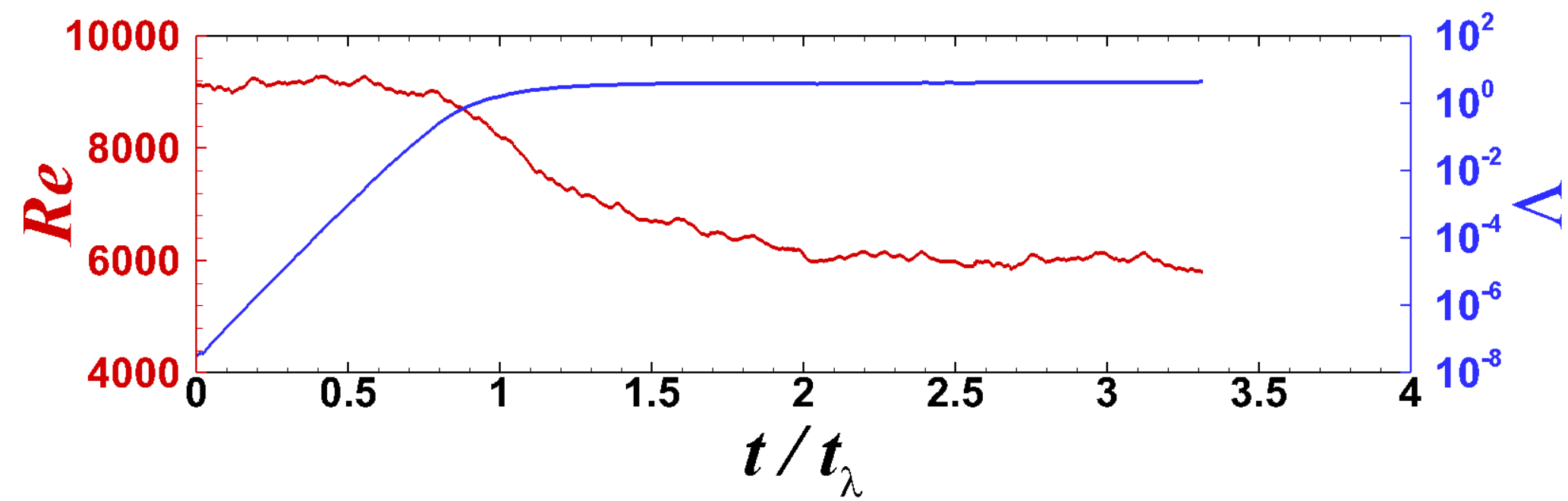}\\
(b)\includegraphics[width=0.5\textwidth,trim={0cm 5.5cm 0cm 7cm},clip]{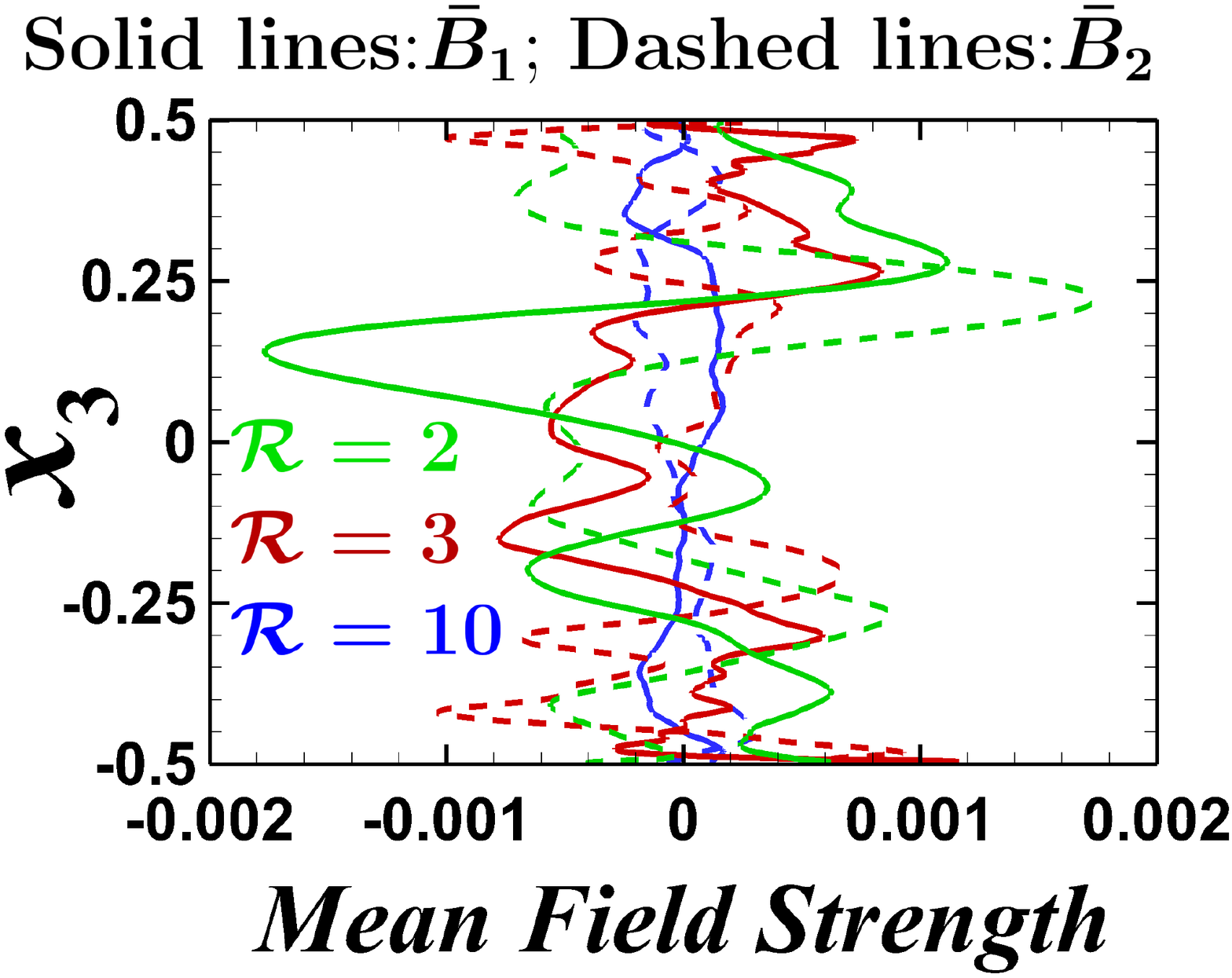}
\caption{(a)Time evolution of Reynolds Number and Elsasser Number for DC at $\mathcal{R}=20$ (b) Vertical variation of mean magnetic fields for different $\mathcal{R}$.}
\label{fig:ic}
\end{figure*}

We start the DC simulations at $\mathcal{R}=20$, by introducing a small magnetic perturbation in a statistically steady RC simulation, as shown in figure \ref{fig:ic}a. The magnetic field shows an exponential growth over approximately one free decay time scale ($t_\lambda=d^2/\lambda$) before it saturates. The decrease in Reynolds Number ($Re=Re_m/Pr_m$), signify decrease in turbulence due to dynamo action at $\mathcal{R}=20$. Simulations at lower values of $\mathcal{R}$ are started using the data of higher $\mathcal{R}$ as the initial condition to save computational time. The mean magnetic field is calculated to outline the behavior of the dynamos, by averaging the field over horizontal planes. This horizontal averaging of any quantity $\phi$ is denoted by over-bar in the following discussion, with $\Bar{\phi}$ denoting the mean and $\phi_{rms}$ representing the \textit{r.m.s.} field. The vertical variation of the mean horizontal magnetic fields ($\Bar{B}_{1}\  \text{and}\  \Bar{B}_{2}$) is depicted in figure \ref{fig:ic}b. It should be noted here that the vertical component of the mean magnetic field ($\Bar{B}_{3}$) is zero everywhere by the definition of averages and solenoidal field conditions. Thus the mean-field remains horizontal. However, as we show later in figure \ref{fig:dynamo}b, the vertical component of the \textit{r.m.s} field (${b}_{3,rms}$) is non-zero except at the walls. We observe one order of magnitude drop in the mean-field strength (figure \ref{fig:ic}b) with increasing $\mathcal{R}$ signifying a transition from large scale to small scale dynamo action.  The magnetic Reynolds number (based on the horizontal scale of convective cells,   \citep{tilgner_2012}) varies as $Re_mE^{1/3}\approx 7-76$ in the range $\mathcal{R}=2-20$ in our simulations. A similar transition in dynamos is reported to occur at $Re_mE^{1/3}=13.5$ \citep{tilgner_2012,tilgner_2014} which lies between $\mathcal{R}=2.5-3 $ our simulations.\\

\begin{table}
  \begin{center}
\def~{\hphantom{0}}
  \begin{tabular}{ccccccc}
       $\mathcal{R}$ & \qquad $Ra$ & \qquad  $Re_m$  & \qquad $\frac{Nu}{Nu_0}$ & \qquad  $\Lambda$ & \qquad  $\Lambda_T$ & \qquad  $Ro_T$ \\[3pt]
       2   & \qquad $3.8300\times10^{9}$ & \qquad 729 & \qquad 0.7466 & \qquad 0.2186 & \qquad 0.0046 & \qquad 0.0220\\
       2.5   & \qquad $4.7875\times10^{9}$ & \qquad 1327 & \qquad 1.2550 & \qquad 0.5163 & \qquad 0.2921 & \qquad 0.0484 \\
       3   & \qquad $5.7450\times10^{9}$ & \qquad 2012 & \qquad 1.7205 &\qquad 1.0313 & \qquad 0.6170 & \qquad 0.0727 \\
       4   & \qquad $7.6600\times10^{9}$ & \qquad 2284 & \qquad 1.4941 &\qquad 1.6834 & \qquad 0.5545 & \qquad 0.1211 \\
       5   & \qquad $9.5750\times10^{9}$ & \qquad 2660 & \qquad 1.4282 &\qquad 2.3994 & \qquad 0.4616 & \qquad 0.1535 \\ 
       10   & \qquad $1.9150\times10^{10}$ & \qquad 4203 & \qquad 1.1714 &\qquad 3.7428 & \qquad 0.2137 & \qquad 0.3539 \\
       20   & \qquad $3.8300\times10^{10}$ & \qquad 7642 & \qquad 1.1377 &\qquad 13.2865 & \qquad 0.1672 & \qquad 0.8663 \\       
  \end{tabular}
  \caption{Statistics of the DC simulations with no-slip boundary conditions at $E=10^{-6}$, $Pr=Pr_m=1$}
  \label{tab:volavg}
  \end{center}
\end{table}

In the following discussion, the DC and RC simulations are compared in terms of their heat transfer, and turbulence properties. Heat transfer is represented by the Nusselt number($Nu$), defined as the total heat to conductive heat transferred from the bottom plate to the top plate. The intensity of turbulence is characterized by the turbulent kinetic energy ($K$), whereas the viscous and the Joule dissipation signifies the conversion of kinetic, and magnetic energy to internal energy via the action of viscous and magnetic diffusion, respectively.

\begin{equation}\label{eqn:nu}
\begin{split}
    Nu=\frac{qd}{k\Delta T}=1+\sqrt{RaPr}\langle u_3\theta\rangle, \hspace{1cm} \langle K \rangle=\langle{\frac{1}{2}u_{i}u_{i}}\rangle \\
       \langle \epsilon _{v} \rangle=\sqrt{\frac{Pr}{Ra}}\Bigg\langle\frac{\partial u_{i}}{\partial x_j}\frac{\partial u_{i}}{\partial x_j}\Bigg\rangle, \hspace{1cm}        \langle \epsilon _{j} \rangle=\sqrt{\frac{Pr}{Ra}}\frac{1}{Pr_m}\Bigg\langle\frac{\partial b_{i}}{\partial x_j}\frac{\partial b_{i}}{\partial x_j}\Bigg\rangle \\
\end{split}
\end{equation}

Here $\langle . \rangle$ denote average over the entire volume. The volume-averaged total heat flux and vertical buoyant energy-flux are denoted by $q$ and $\langle u_3\theta\rangle$. Subscript $"0"$ is used to represent the properties without magnetic field (RC cases) in the rest of this paper. 

\section{Result}
\begin{figure}
\centering
\includegraphics[width=0.5\linewidth,trim={0 6.15cm 0 6.15cm},clip]{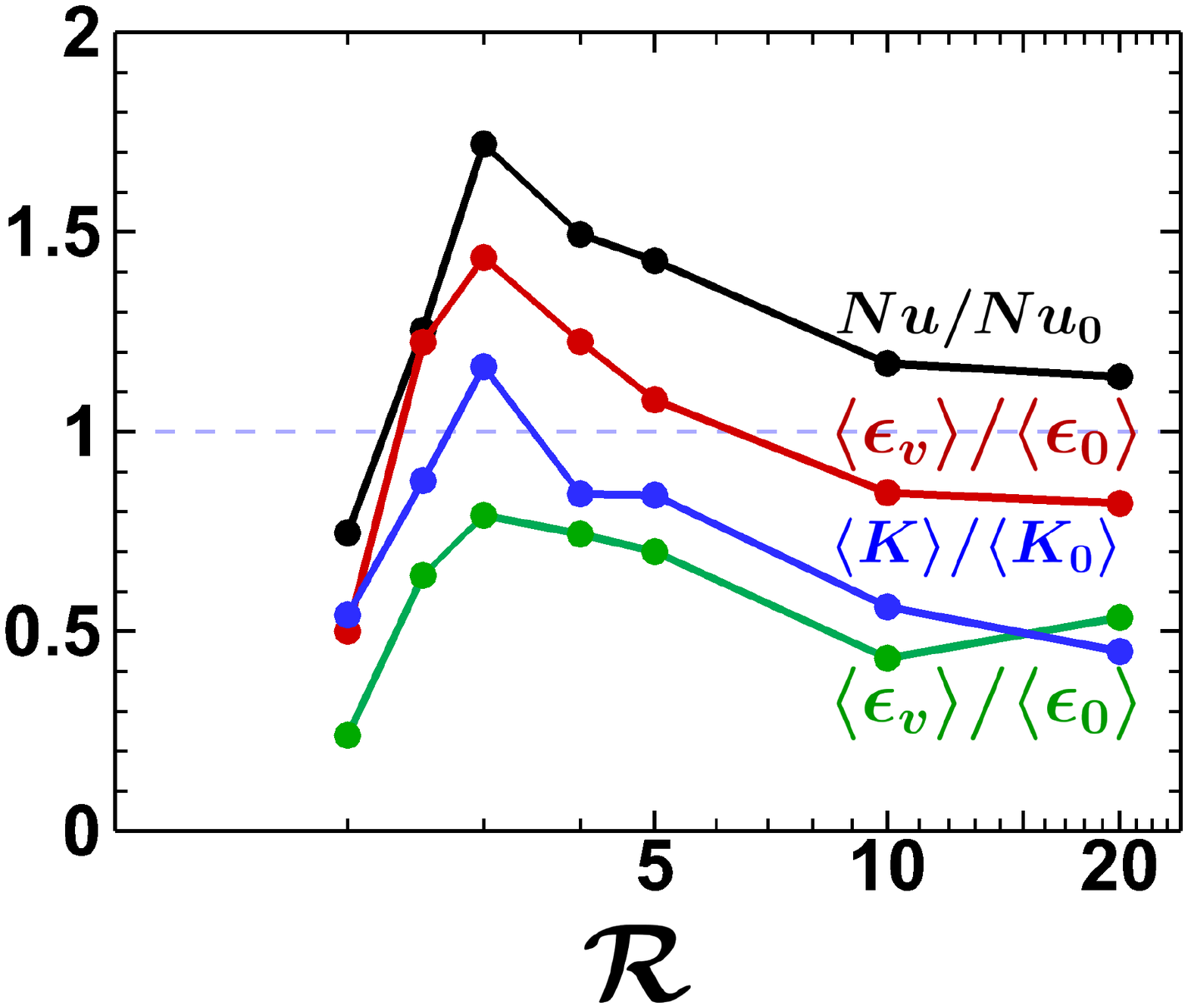}
\caption{Ratio of Nusselt Number, Turbulent Kinetic Energy, viscous dissipation, and Joule dissipation as a function of convective supercriticality. Subscript $"0"$ is used to represent non-magnetic simulation.}
\label{fig:nusselt}
\end{figure}

We begin the analysis by comparing our simulation results for DC and RC in terms of their global heat transfer and turbulence properties. In figure \ref{fig:nusselt}, the ratio of the Nusselt number for DC to RC ($Nu/Nu_{0}$) is plotted with respect to the convective supercriticality ($\mathcal{R}$). The dynamo action associated with the magnetic field results in increased heat transfer for DC, except at $\mathcal{R}=2$.  Maximum enhancement occurs at $\mathcal{R}=3$ where the heat transfer of DC is more than $70\%$ as compared to RC. The experiment in a rotating cylinder by \citet{king_2015} and the geodynamo simulation in a spherical shell model by \citet{Yadav_2016} have reported a similar heat transfer enhancement peak. However, this optimum heat transfer enhancement owing to dynamo action is a novel finding for a plane layer dynamo. Interestingly, the peak in heat transfer is found to occur at $\Lambda \approx 1$ (see table \ref{tab:volavg}), similar to the RMC experiments of \citet{king_2015} and the geodynamo simulation of \cite{Yadav_2016}. When the Lorentz force is of similar magnitude with the Coriolis force, the global magnetorelaxation process should enhance heat transport in RMC \citep{king_2015}. However, this leading order magnetostrophic balance for $\Lambda=O(1)$ may not represent the primary balance in DC \citep{calkins_2018,soderlund_2015,aurnou_2017}. The traditional definition of Elsasser number($\Lambda=\sigma B_\textit{r.m.s}^2/\rho\Omega$) gives a correct ratio of the Lorentz and the Coriolis forces only for small $Re_m\leq O(1)$, which is not valid for a dynamo \citep{soderlund_2015,aurnou_2017}. As the global force balance reported in the simulation \citep{Yadav_2016} is geostrophic, not magnetostrophic, a similar magnetorelaxation process as reported in the RMC experiment of \citep{king_2015}, may not be the reason for heat transfer enhancement in the DC simulations of \citep{Yadav_2016}. The present study indicates a local increase in the Lorentz force near the top and bottom plates signifying the importance of the thermal boundary layer dynamics for the heat transfer enhancement in dynamo simulations, instead of any global force balance.\\

The global statistics of the underlying turbulent motion are shown in figure \ref{fig:nusselt} to understand the heat transfer enhancement phenomenon. We observe that the ratio of the turbulent kinetic energy ({\it{t.k.e.}}) of the DC simulations to the RC simulations, $\langle K \rangle/\langle K_{0} \rangle$, achieves a maximum at $\mathcal{R}=3$ indicating higher turbulence intensity in dynamo convection as compared to rotating convection. We further confirm this enhanced turbulence activity by showing the variation of the ratio of viscous dissipation,$\langle\epsilon_{v}\rangle/\langle \epsilon_{0}\rangle$, in figure \ref{fig:nusselt}. A similar peak in dissipation ratio for $\mathcal{R}=3$ is observed. Additionally, the part of kinetic energy converted into magnetic energy dissipates via Joule dissipation. The normalized Joule dissipation of magnetic energy, $\langle\epsilon_{j}\rangle/\langle \epsilon_{0}\rangle$, manifests a similar peak. These observations confirm increased turbulence for $\mathcal{R} = 3$ that promotes mixing and scalar transport, resulting in heat transfer enhancement.\\

\begin{figure*}
\centering
(a) \includegraphics[width=0.4885\linewidth,trim={0.5cm 7.6cm 0cm 8cm},clip]{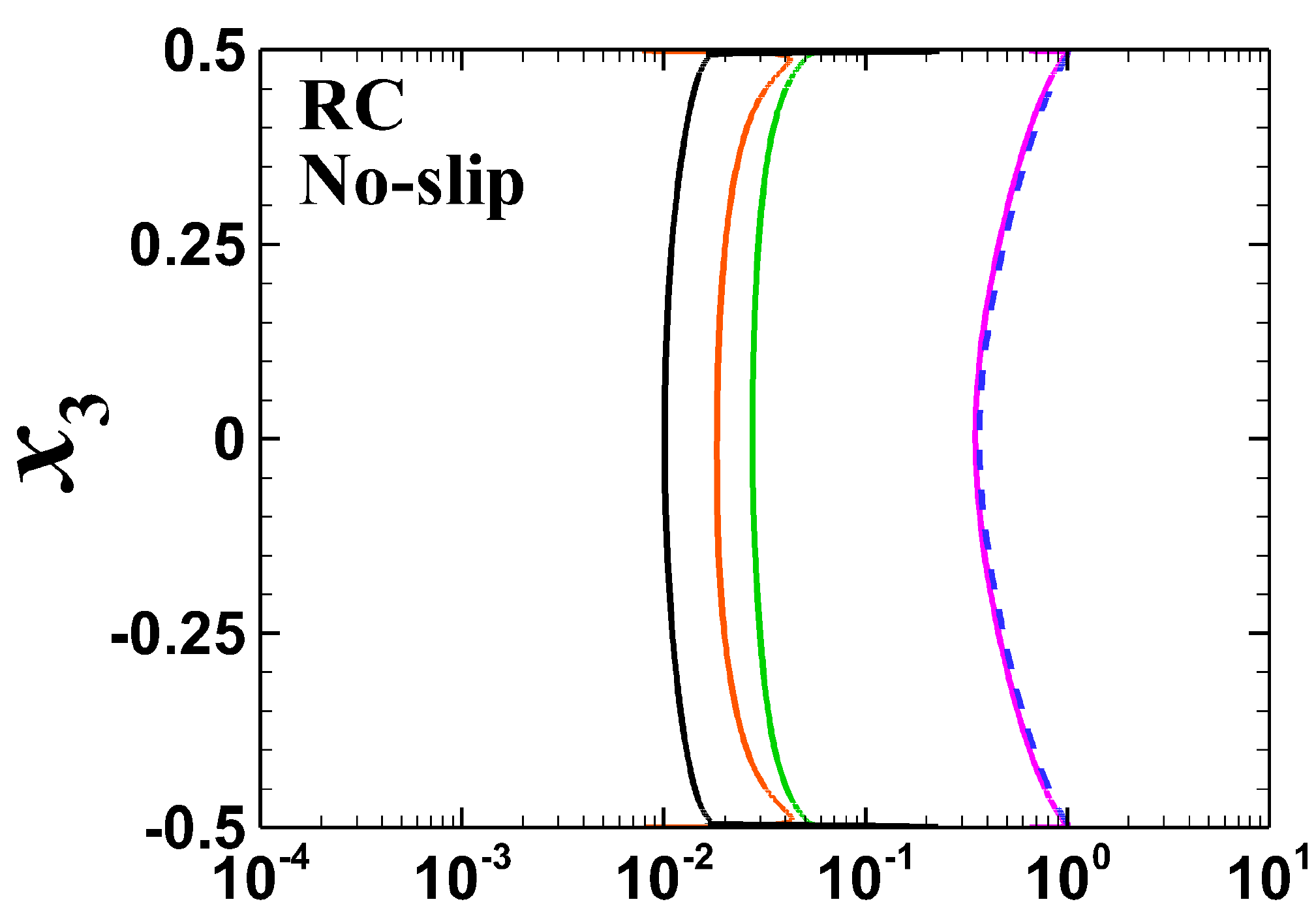}
(b) \includegraphics[width=0.4425\linewidth,trim={2.5cm 7.6cm 0cm 8cm},clip]{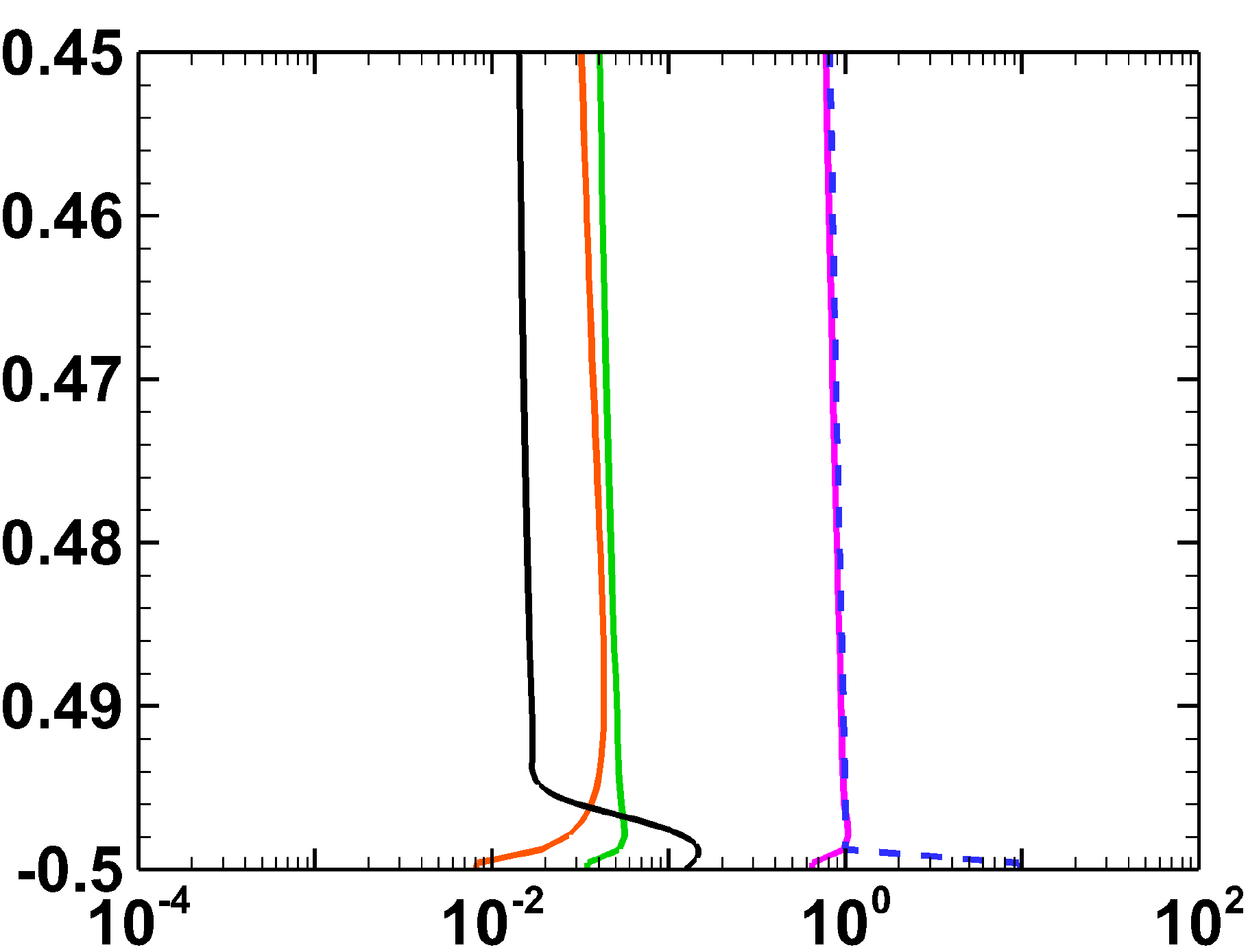}\\
(c) \includegraphics[width=0.4885\linewidth,trim={0.5cm 7.6cm 0cm 8cm},clip]{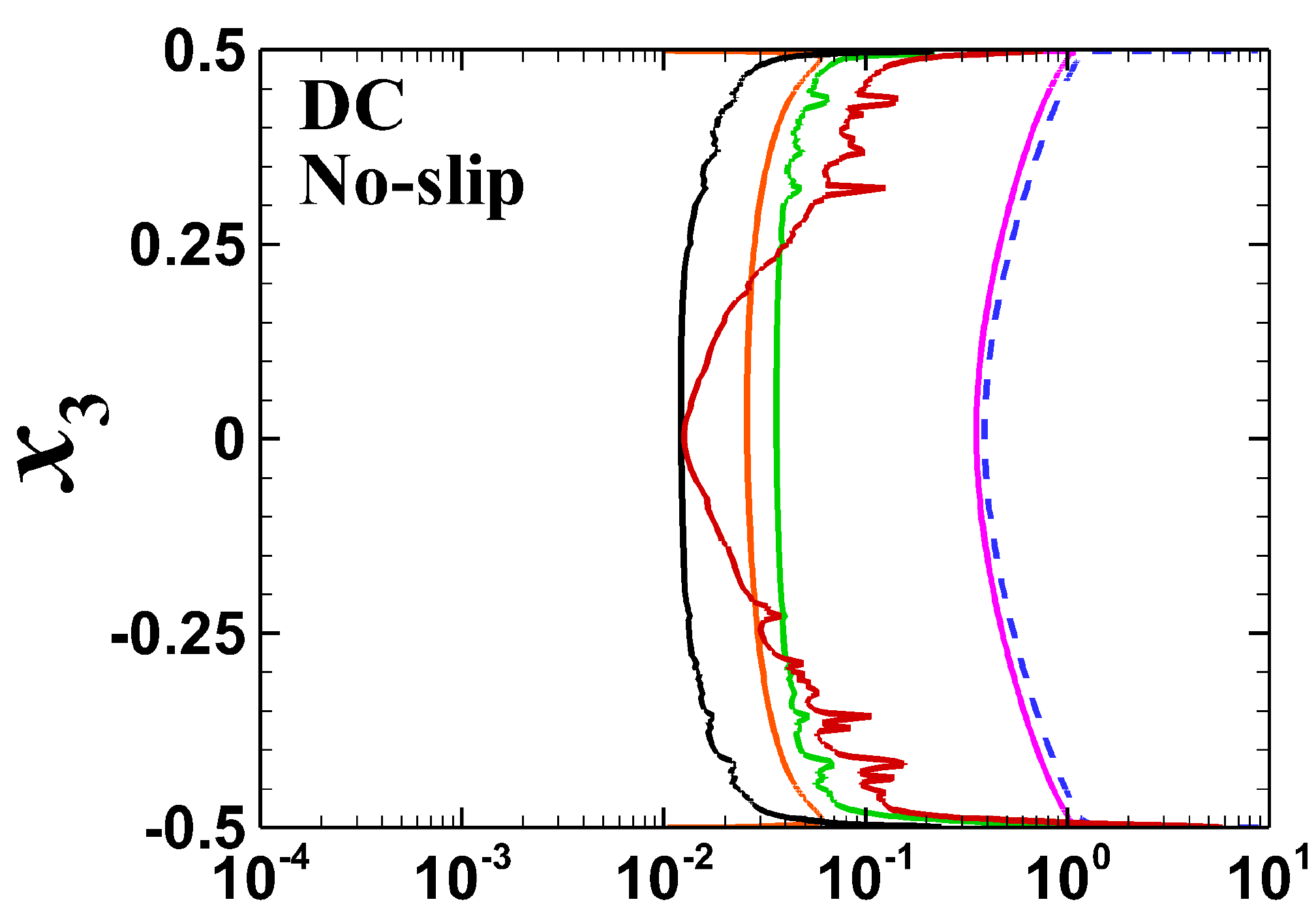}
(d) \includegraphics[width=0.4425\linewidth,trim={2.5cm 7.6cm 0cm 8cm},clip]{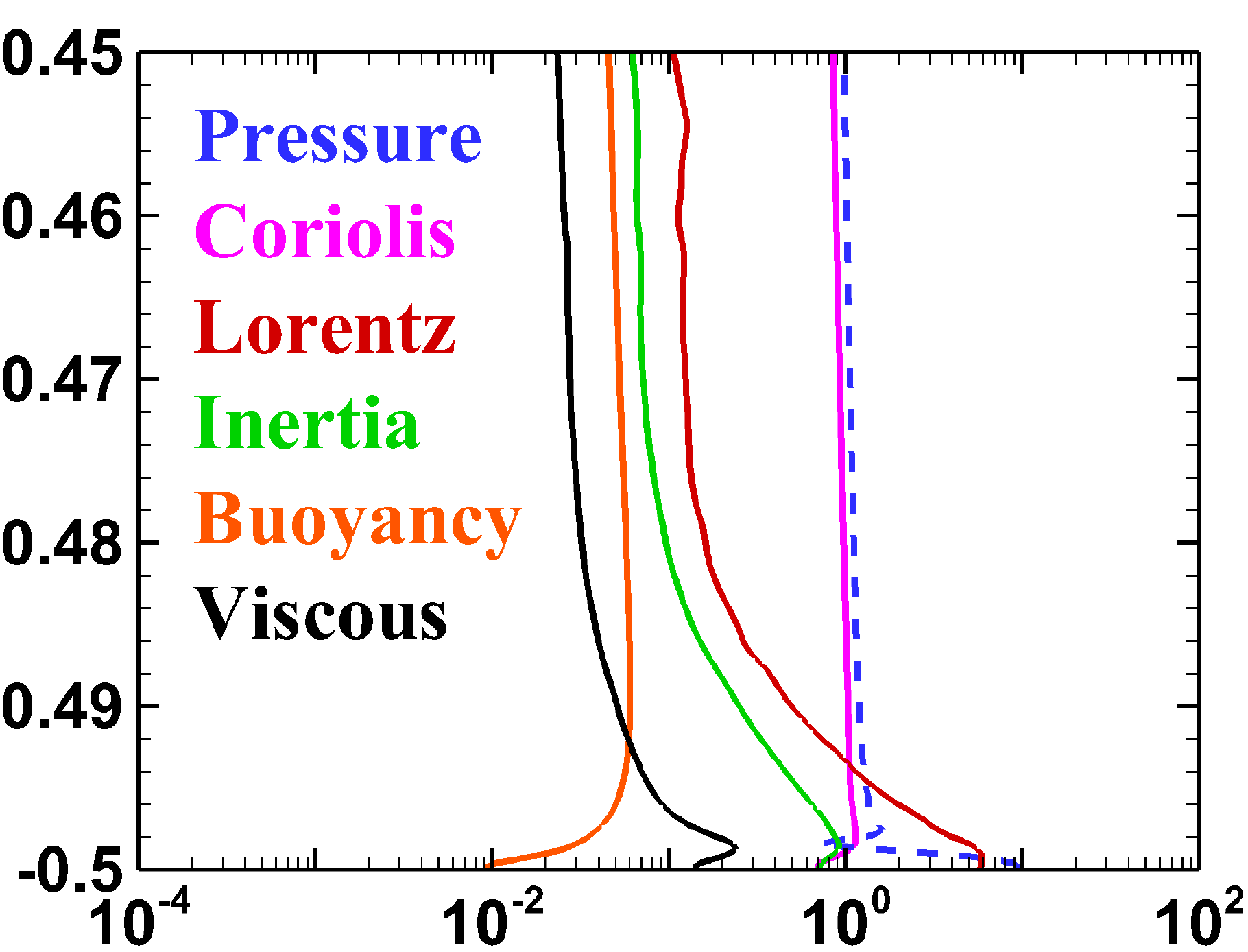}\\
(e) \includegraphics[width=0.4885\linewidth,trim={0.5cm 6cm 0cm 8cm},clip]{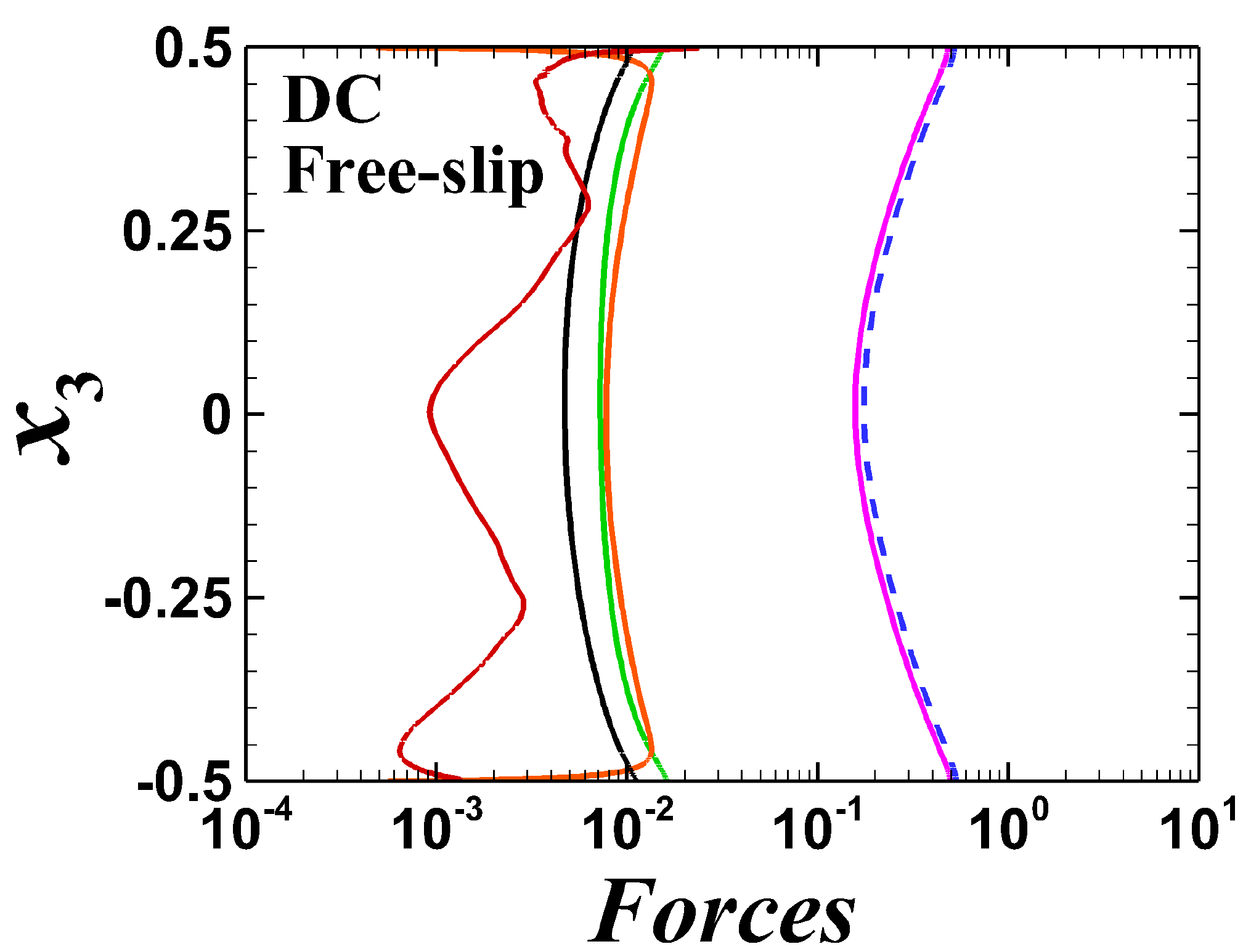}
(f) \includegraphics[width=0.4425\linewidth,trim={2.5cm 6cm 0cm 8cm},clip]{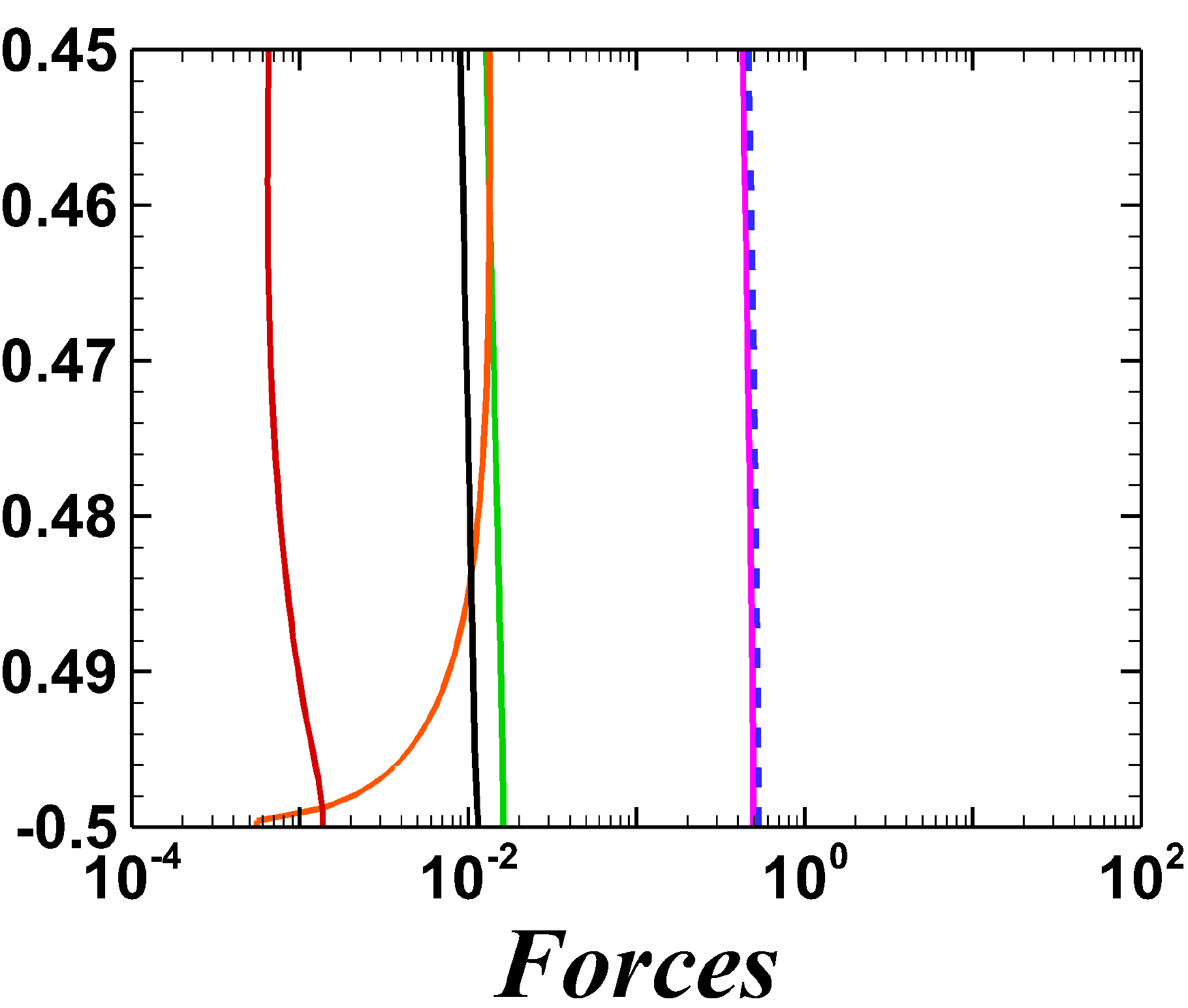}
\caption{Vertical variation of forces at $\mathcal{R}=3$ for (\textit{a}, \textit{b}) no-slip RC, (\textit{c}, \textit{d}) no-slip DC and (\textit{e}, \textit{f}) free-slip DC. The horizontally averaged force distribution is shown in the bulk(left column) and near the bottom plate(right column).}
\label{fig:forces}
\end{figure*}

To explore the possible reasons for this increased turbulence and heat transfer enhancement, we can compare the vertical variation of forces in figure \ref{fig:forces} for DC and RC with both no-slip and free-slip boundary conditions at $\mathcal{R}=3$. The forces are calculated from the \textit{r.m.s} values of the terms in momentum equation (\ref{eqn:momentum_nd}) by averaging over horizontal planes \citep{yan_2021}. The leading order geostrophic balance between pressure and the Coriolis forces are apparent for all cases as shown in figures \ref{fig:forces} a, c and e. Geostrophic balance is known to be the characteristic of strongly rotating convection \citep{king_2012}. Small departures from this balance due to the other forces make quasi-geostrophic convection (even turbulence) possible in such flows, making these other forces important for the dynamics. In the present study, buoyancy, viscous, and inertia constitute the quasi-geostrophic force balance in RC at $\mathcal{R}=3$ (see figure \ref{fig:forces}a). Although the viscous diffusion term is smallest in the bulk, it increases near the Ekman layer at the bottom plate as shown in figure \ref{fig:forces}b and dominates the quasi-geostrophic balance. For the no-slip boundary condition, the velocities are zero at the wall, which results in an increase in the pressure force at the walls (figure \ref{fig:forces}b,d). In figure \ref{fig:forces}c, the Lorentz force, which is at least one order of magnitude lower than Coriolis force in the bulk, decreases from the wall towards the center of the domain, reaching a minimum. Therefore, the global magnetostrophic balance between the Lorentz and the Coriolis forces, as found in the previous RMC experiment \citep{king_2015}, may not be the reason for the heat transfer enhancement in the present DC simulations. A closer look near the bottom plate for DC at $\mathcal{R}=3$ reveals a different dynamical behavior in figure \ref{fig:forces}d. Here, the Lorentz force increases nearly three orders of magnitude from its value at the center of the domain to become the dominant force at the wall. This increase in the Lorentz force near the wall can attenuate the turbulence suppressing effect of the Coriolis force resulting in enhanced turbulence, mixing, and heat transfer. We also observe a similar increase in Lorentz force near the boundary for $\mathcal{R}=2.5,4 \ \text{and}\ 5$ (figures not shown). Apart from a dissimilar vertical variation, the Lorentz force for the free-slip boundary condition case (figure \ref{fig:forces}e) is one order of magnitude lower compared to the no-slip boundary condition case (figure \ref{fig:forces}c). Furthermore, no enhancement in the Lorentz force is observed for the free-slip boundary condition case near the wall (see figure \ref{fig:forces}f).\\

\begin{figure*}
\centering
(a) \includegraphics[width=0.489\linewidth,trim={1cm 6cm 0cm 8cm},clip]{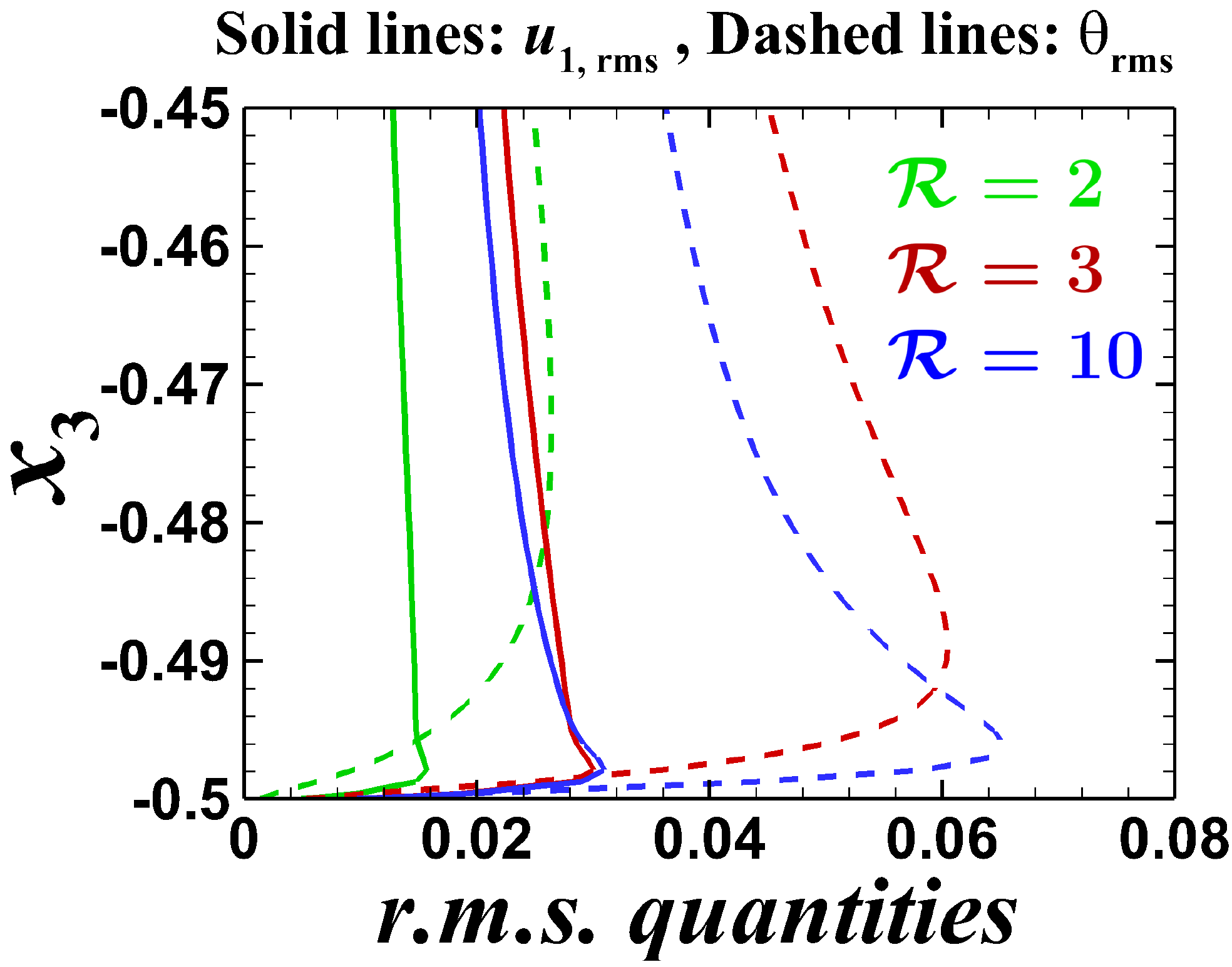}
(b) \includegraphics[width=0.44\linewidth,trim={3cm 6cm 0cm 9cm},clip]{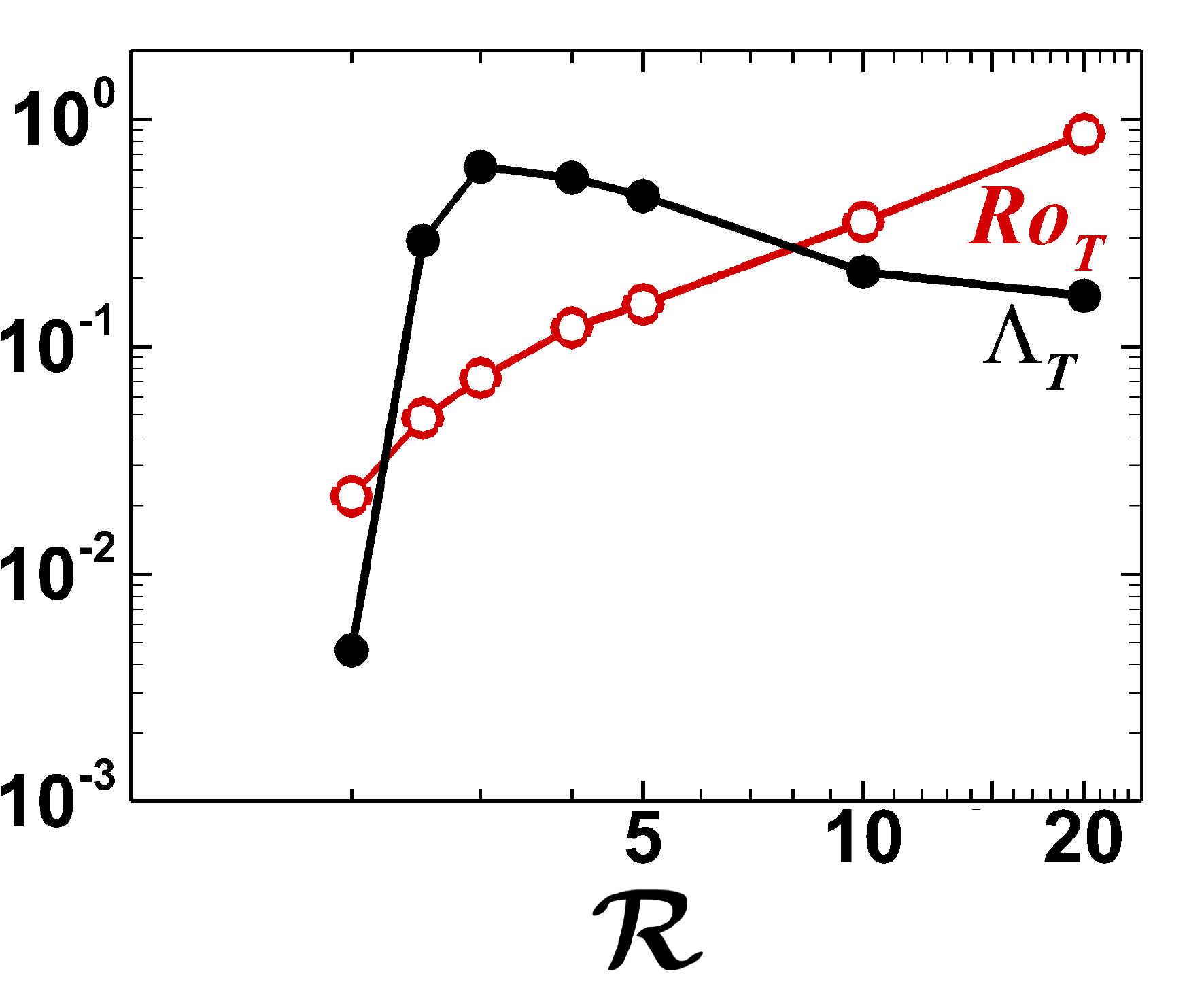}\\
\caption{(\textit{a}) Horizontal \textit{r.m.s} velocity(solid lines) and temperature profiles(dashed lines) near wall for DC cases, (\textit{b}) variation of local Elsasser and the Rossby number at the thermal boundary layer edge with convective supercriticality.}
\label{fig:trms}
\end{figure*}

Further analysis on the boundary layer characteristics can be made to assess the relative magnitudes of the Lorentz and the Coriolis forces near the walls. The variation of the \textit{r.m.s.} horizontal velocity ($u_{1,r.m.s}$), and the \textit{r.m.s.} temperature ($\theta_{r.m.s}$) in the vertical direction near the bottom plate is shown in figure \ref{fig:trms}a for DC cases at $\mathcal{R}=2,3\ \text{and}\ 10$. Here, we can define the Ekman boundary layer (EBL) and thermal boundary layer (TBL) as the regions adjacent to the plates where viscous, and thermal diffusion are important. In strongly rotating convection, heat is transported by columnar structures, and plumes that are formed due to the instability of the TBL \citep{king_2012}. The heat transfer characteristics of such systems depend on the relative thickness of the two layers \citep{king_2009,king_2012}. The distance of the peaks in \textit{r.m.s.} horizontal velocity, and temperature  from the bottom boundary can be used to estimate the EBL and TBL thickness ($\delta_{E}$ and $\delta_{T}$) respectively from figure \ref{fig:trms}a \citep{king_2009}. If $\delta_{T}>\delta_{E}$, the strong Coriolis force stabilizes the thermal layer, impeding the formation of plumes and thereby restricting heat transfer \citep{king_2012}. The thermal boundary layer thickness is higher than the Ekman layer for the DC cases as demonstrated in figure \ref{fig:trms}a. Similar observations are true for RC cases (figure not shown) indicating a strong stabilizing effect of Coriolis force in all simulations. To determine the effect of the Lorentz force in DC cases, we define a local Elsasser number($\Lambda_{T}$) at the edge of TBL that measures the ratio of the Lorentz and the Coriolis forces at this location. We emphasize here that this Elsasser number is computed directly from the \textit{r.m.s.} magnitudes of the forces (see figure \ref{fig:forces}), not from the traditional definition($\Lambda$). The variation of $\Lambda_{T}$ with $\mathcal{R}$ at the TBL edge near the bottom boundary is shown in figure \ref{fig:trms}b. The two forces can be seen to be of the same order of magnitude near the edge of the TBL for $\mathcal{R}=2.5,3,4 \ \text{and}\ 5$. The local Elsasser number also achieves a maximum at $\mathcal{R}=3$, similar to the Nusselt number ratio (see figure \ref{fig:nusselt}), indicating a correlation between the increase in the Lorentz force and the heat transport behavior. The increase in the Lorentz force at the thermal boundary layer effectively mitigates the stabilizing effect of the Coriolis force and can be seen as a local magnetorelaxation of the Taylor-Proudman constraint, resulting in enhanced turbulence and heat transport. Though the inertia force, represented by the Rossby Number ($Ro_T$, the ratio of inertia to Coriolis force) in figure \ref{fig:trms}b, also increases near the wall, the monotonically increasing trend does not correlate with the heat transfer behavior. \\



\begin{figure*}
\centering
(a) \includegraphics[width=0.484\linewidth,trim={1cm 6cm 0cm 8cm},clip]{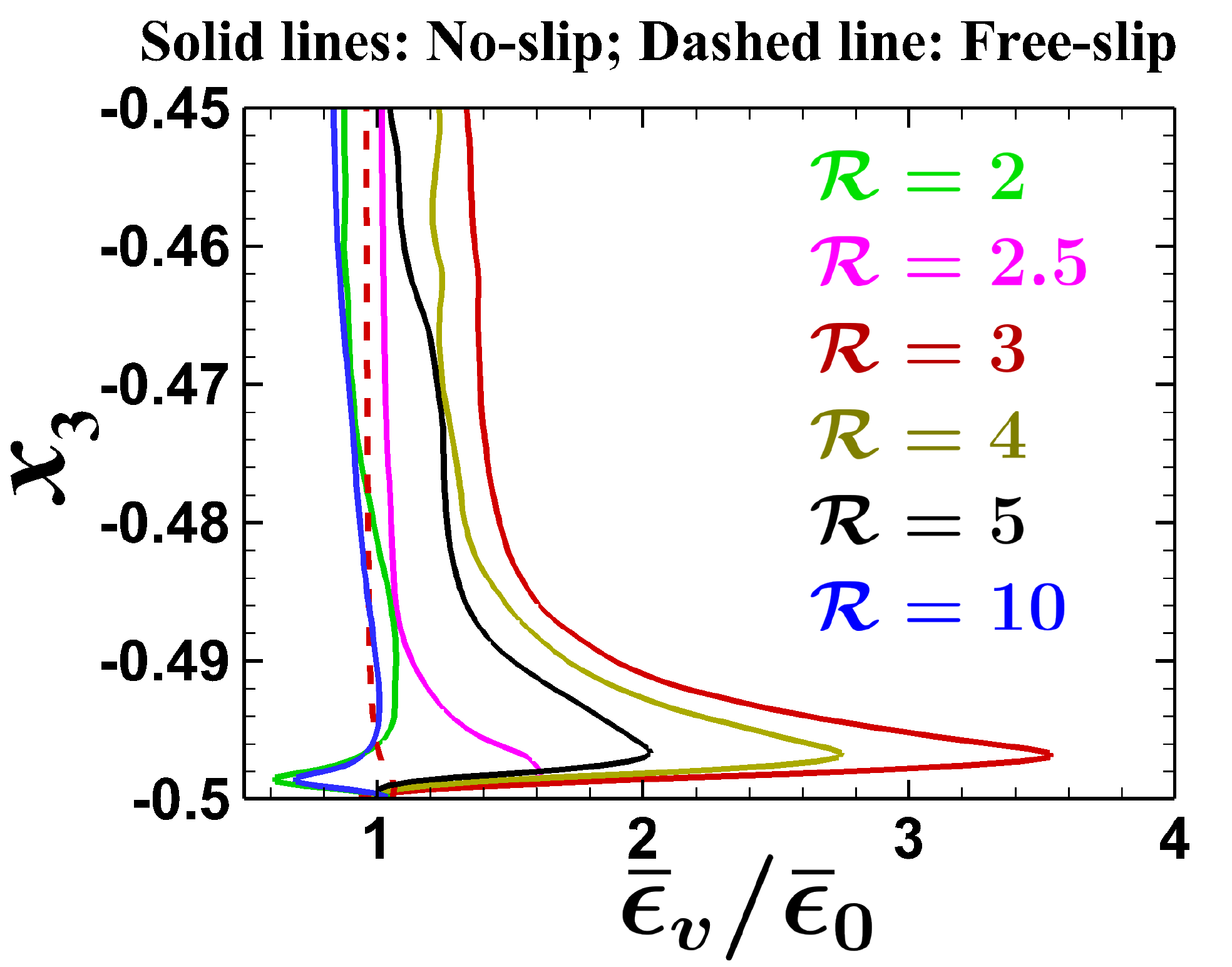}
(b) \includegraphics[width=0.446\linewidth,trim={2.7cm 6cm 0cm 8cm},clip]{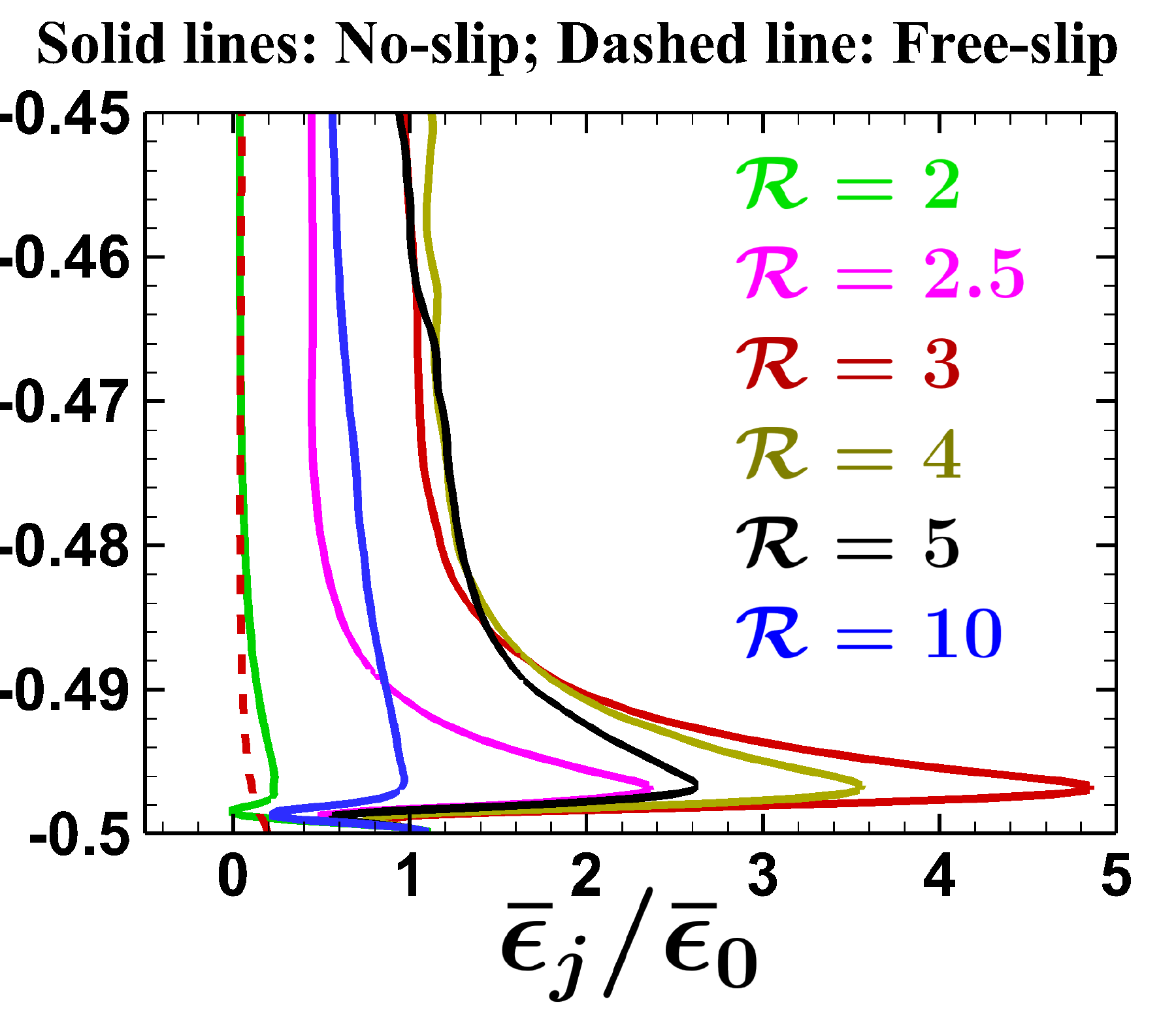}
\caption{ Vertical variation of (\textit{a}) viscous dissipation ratio, and (\textit{b}) Joule dissipation ratio near the bottom plate for DC cases with no-slip (solid lines) and free-slip (dashed line) boundary conditions.}
\label{fig:vjd}
\end{figure*}

To corroborate the local magnetorelaxation process, and enhancement in turbulence, we show the vertical variation of viscous dissipation ratio ($\Bar{\epsilon}_{v}/\Bar{\epsilon}_{0}$) near the bottom wall in figure \ref{fig:vjd}a. This dissipation ratio is computed after averaging the dissipation for DC, and RC simulations over the horizontal directions. We observe an increase in viscous dissipation for $\mathcal{R}=2.5,3,4\ \textrm{and}\ 5$, near the bottom boundary, suggesting an increase in turbulence near the walls because of the local magnetorelaxation. This enhancement in viscous dissipation is highest for $\mathcal{R}=3$ indicating higher turbulence intensity among all the other cases. Similar enhancement in Joule dissipation can also be observed in figure \ref{fig:vjd}b. However, no such enhancement in viscous and Joule dissipation is found at $\mathcal{R} = 3$ for free-slip boundary condition (dashed lines).\\

\begin{figure*}
\centering
(a) \includegraphics[width=0.465\linewidth,trim={0.5cm 3.5cm 0.5cm 4.5cm},clip]{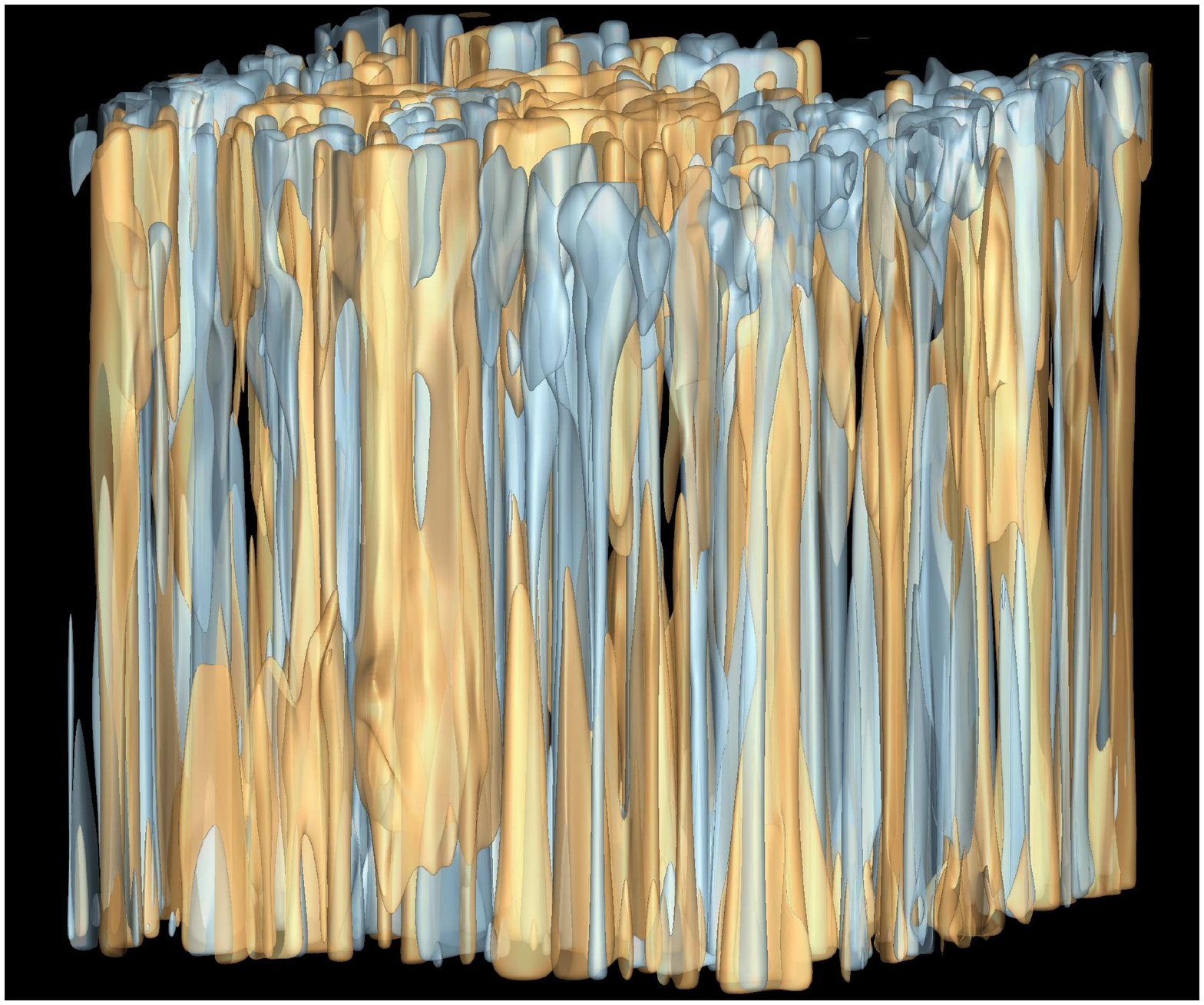}
(b) \includegraphics[width=0.465\linewidth,trim={0.5cm 3.5cm 0.5cm 4.5cm},clip]{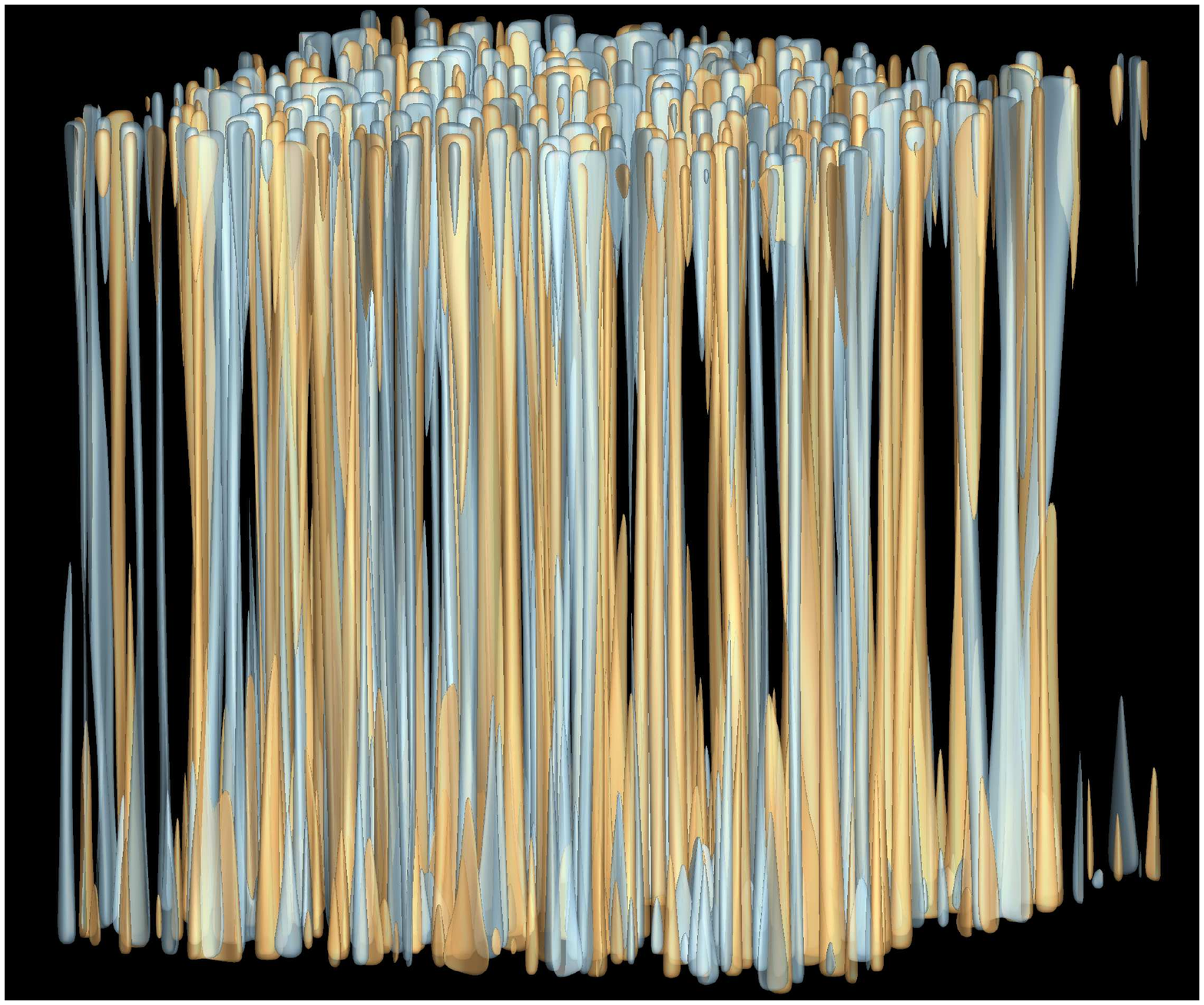}\\
(c) \includegraphics[width=0.465\linewidth,trim={0.5cm 3.5cm 0.5cm 6.5cm},clip]{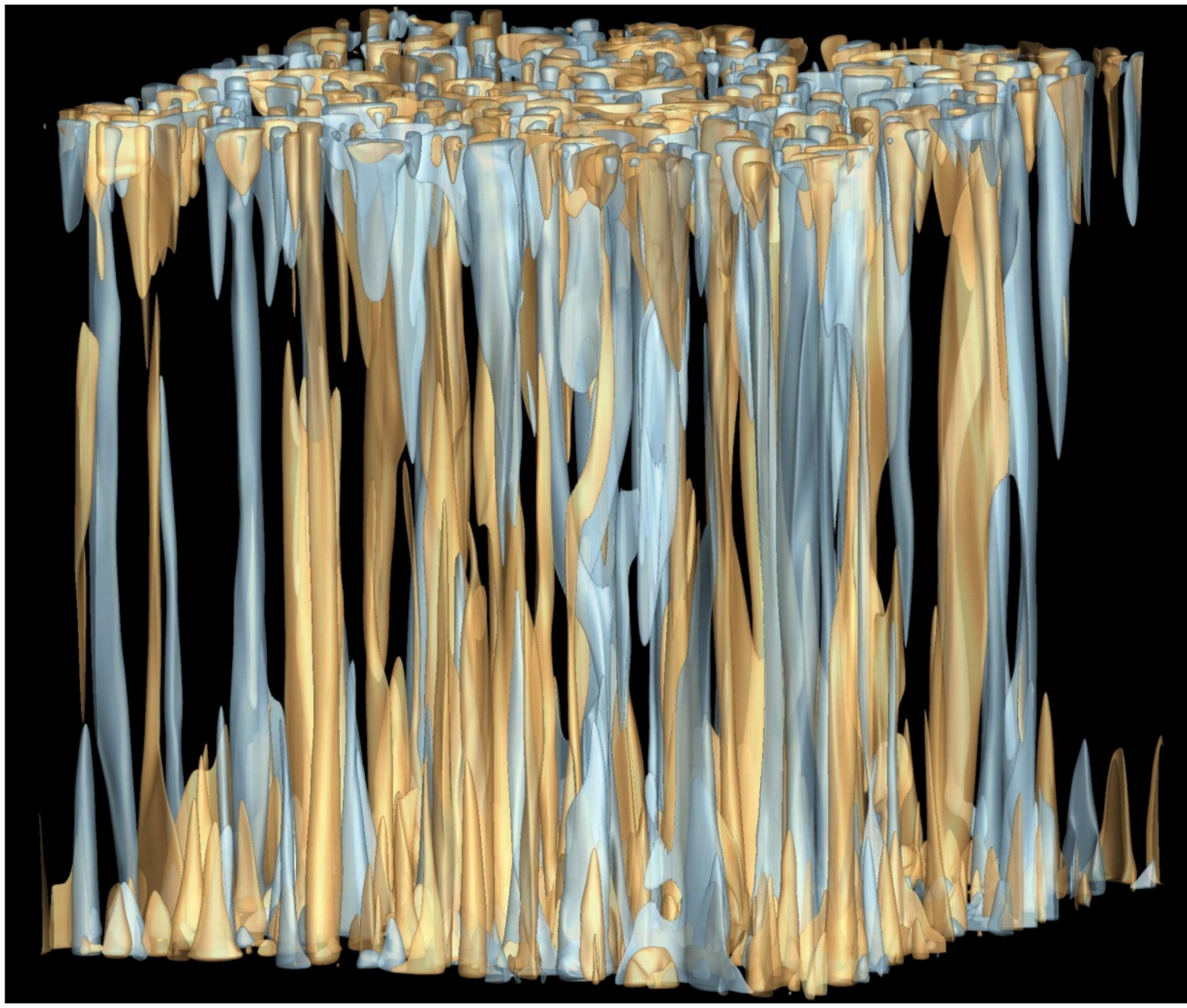}
(d) \includegraphics[width=0.465\linewidth,trim={0.5cm 3.5cm 0.5cm 6.5cm},clip]{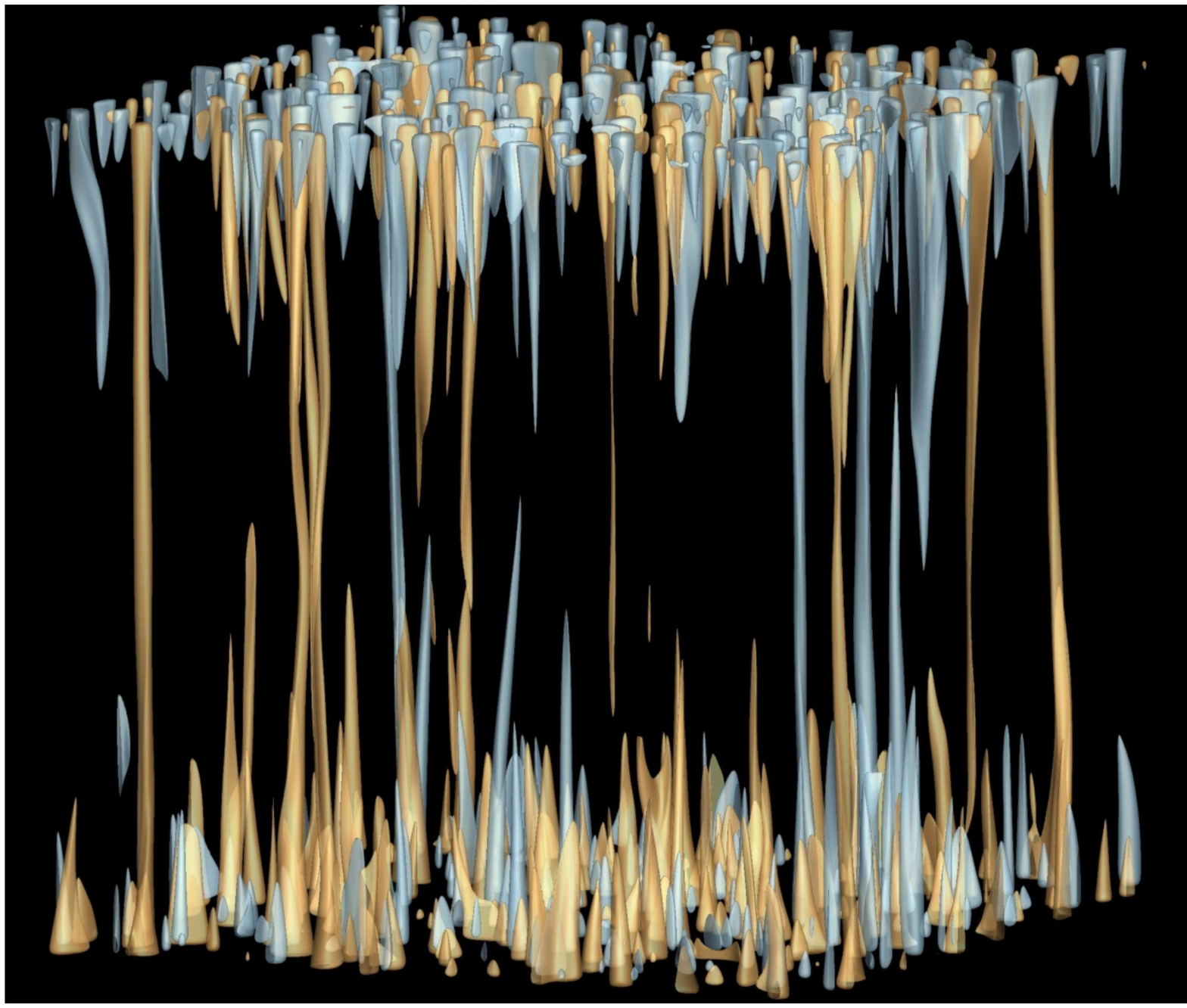}\\
(e) \includegraphics[width=0.465\linewidth,trim={0.5cm 3.5cm 0.5cm 4.5cm},clip]{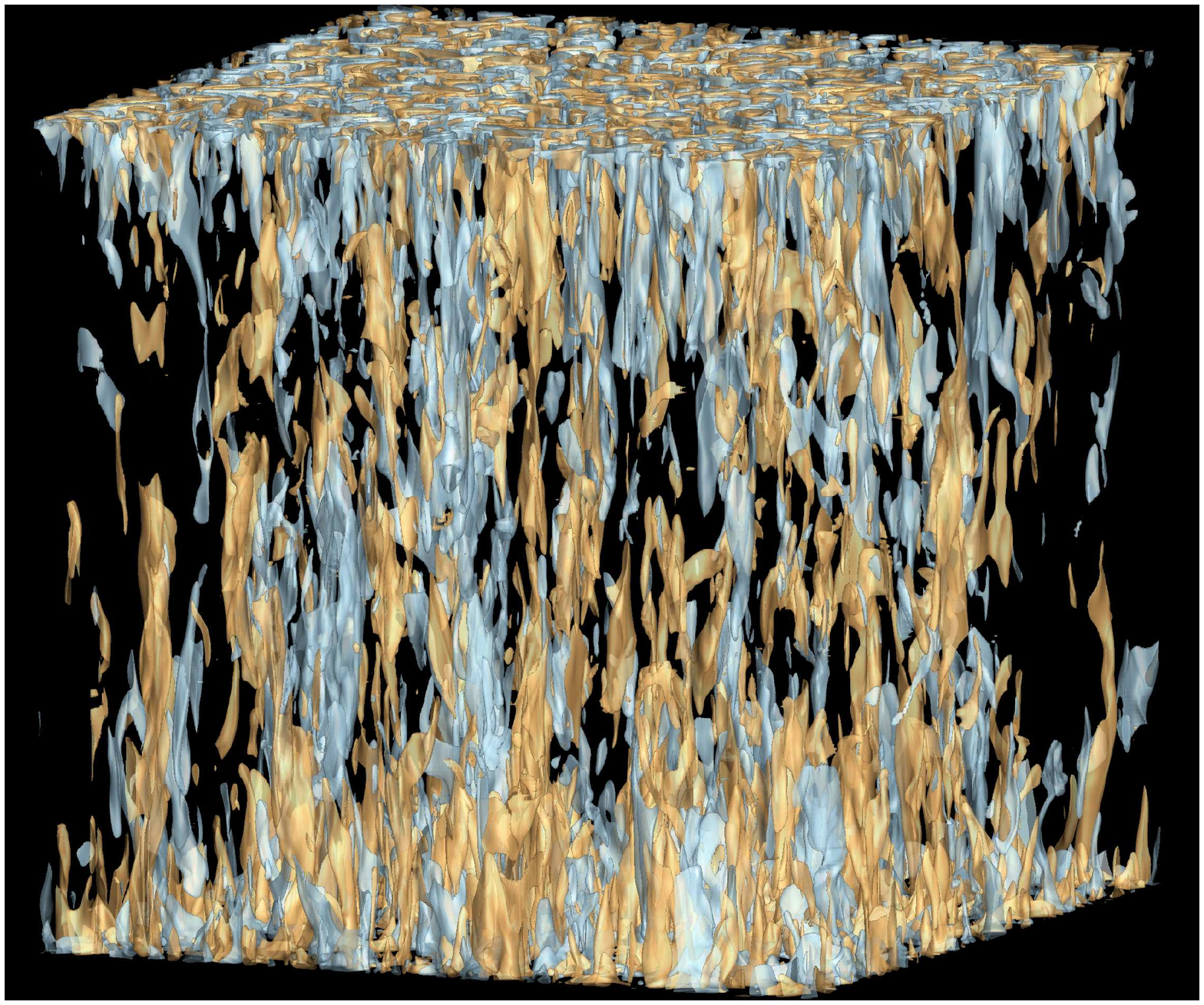}
(f) \includegraphics[width=0.465\linewidth,trim={0.5cm 3.5cm 0.5cm 4.5cm},clip]{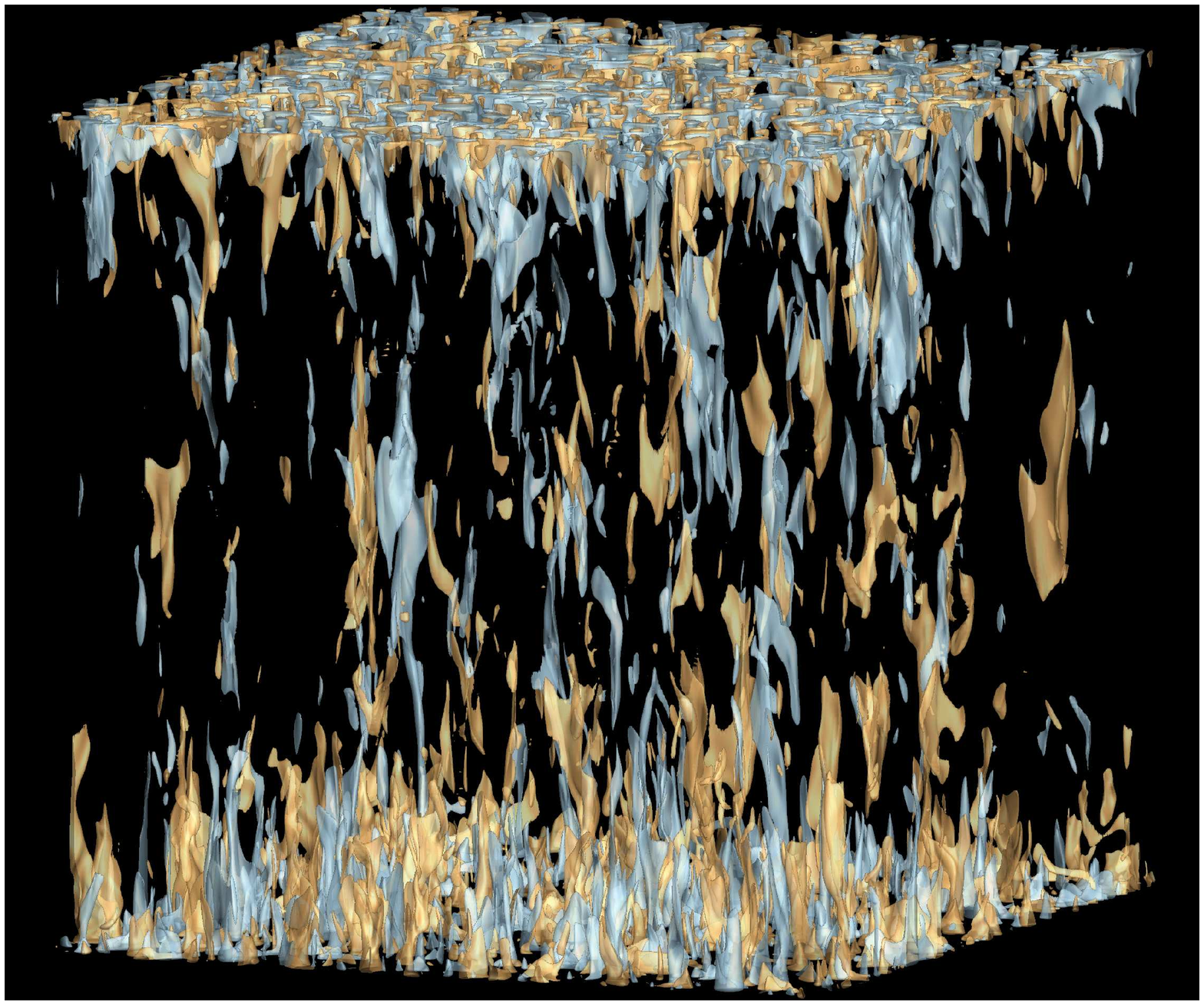}
\caption{Isosurfaces of temperature perturbation for DC (left column) and RC cases (right column). Hot (orange) and cold (grey) columnar and plume structures are visualized by isosurface values $\pm0.03$ for $\mathcal{R}=2$ \textit{(a,b)},  $\pm 0.07$ for $\mathcal{R}=3$ \textit{(c,d)} and $\pm 0.07$ $\mathcal{R}=10$ \textit{(e,f)}.}
\label{fig:rhop}
\end{figure*}

Isosurfaces of temperature perturbation for DC (left column) and RC (right column) are shown in figure \ref{fig:rhop} to provide visualizations of the thermal structures. The horizontally averaged temperature variation is subtracted from the instantaneous temperature field to compute these temperature perturbations. For RC, the flow morphology changes from columnar cells at $\mathcal{R}=2$ (figure \ref{fig:rhop}b) to plumes at $\mathcal{R}=3$ (figure \ref{fig:rhop}d) to geostrophic turbulence dominating at larger thermal forcing at $\mathcal{R}=10$ (figure \ref{fig:rhop}f). The transition of flow morphology from columnar cells to plumes occur at $RaE^{4/3}\sim22$ whereas the transition from plumes to geostrophic turbulence happens at $RaE^{4/3}\sim55$ \citep{julien_2012}. For the present RC simulations at $\mathcal{R}=2,3\ \text{and}\ 10$, the values of $RaE^{4/3}=15.2,22.8\  \text{and}\ 76$, agrees well with the predictions of this transition regime. Thick columnar convection cells with larger horizontal scales are observed for DC at $\mathcal{R}=2$ in figure \ref{fig:rhop}a, due to higher effective viscosity in the presence of a magnetic field \citep{chandrasekhar_1961}. The temperature perturbation field at $\mathcal{R}=3$ constitutes both columnar structures, and plumes as shown in figure \ref{fig:rhop}c and figure \ref{fig:rhop}d. Compared to RC in figure \ref{fig:rhop}d, DC at $\mathcal{R}=3$ (figure \ref{fig:rhop}c) is more populated by columnar structures, and plumes. This increased population of columnar structures and plumes is expected due to the local magnetorelaxation phenomenon near the thermal boundary layer as reported in figures \ref{fig:trms} and \ref{fig:vjd}. The upwards (downwards) motion of the hot(cold) plumes from the bottom(top) wall towards the bulk transport momentum, and thermal energy to increase turbulence and heat transport, as depicted in figure \ref{fig:nusselt}. The horizontal distortion of the thermal structures in figure \ref{fig:rhop}c also indicates the relaxation of the Taylor-Proudman constraint due to the magnetic field. Convection at $\mathcal{R}=10$ occurs in the geostrophic turbulence regime, as apparent from figures \ref{fig:rhop}e,f with more turbulence and abundant small scale structures as compared to lower $\mathcal{R}$ cases.\\

\begin{figure*}
\centering
(a) \includegraphics[width=0.4815\linewidth,trim={0.4cm 3.5cm 0.0cm 5cm},clip]{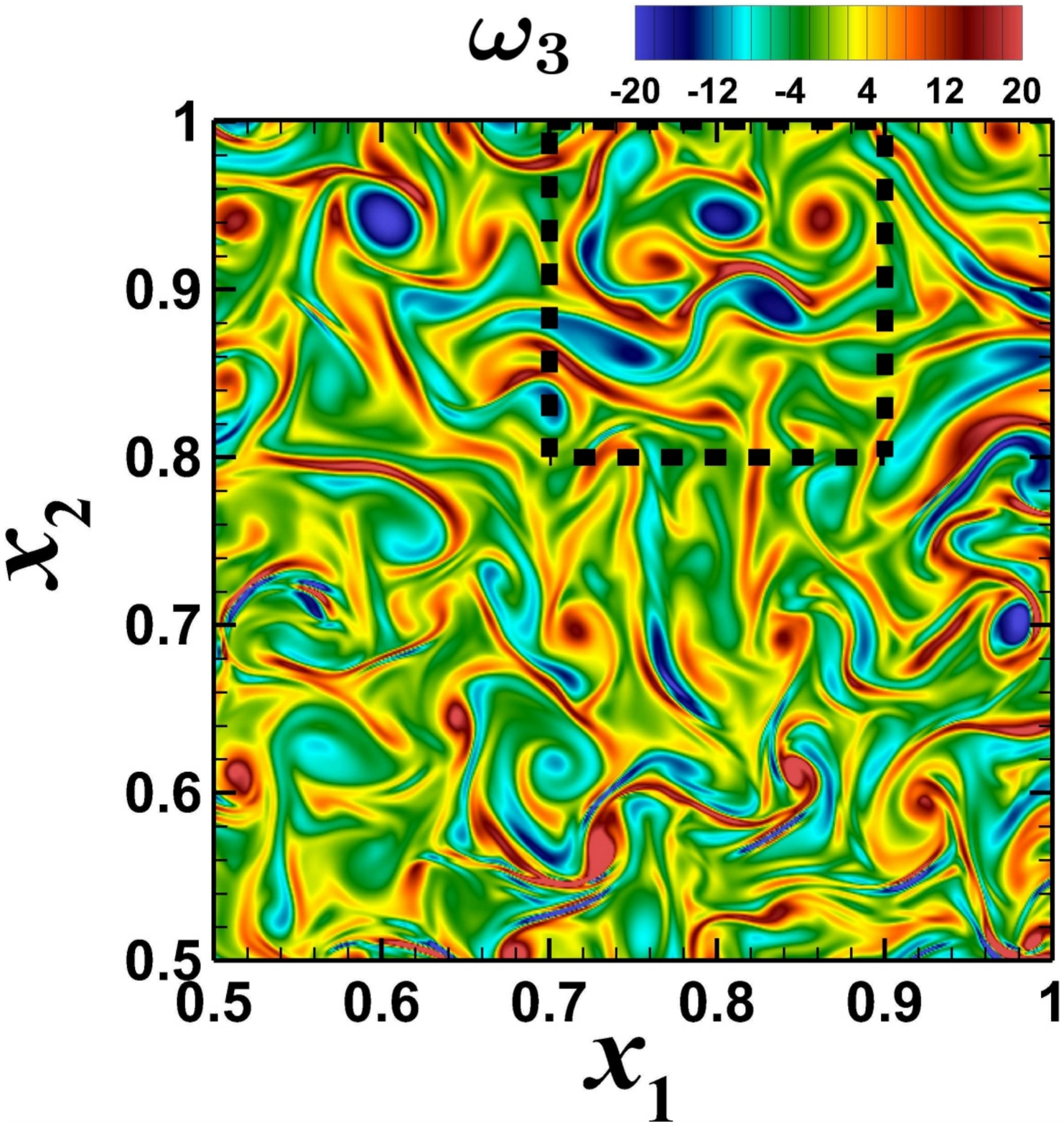}
(b) \includegraphics[width=0.4485\linewidth,trim={2.0cm 3.5cm 0.0cm 5cm},clip]{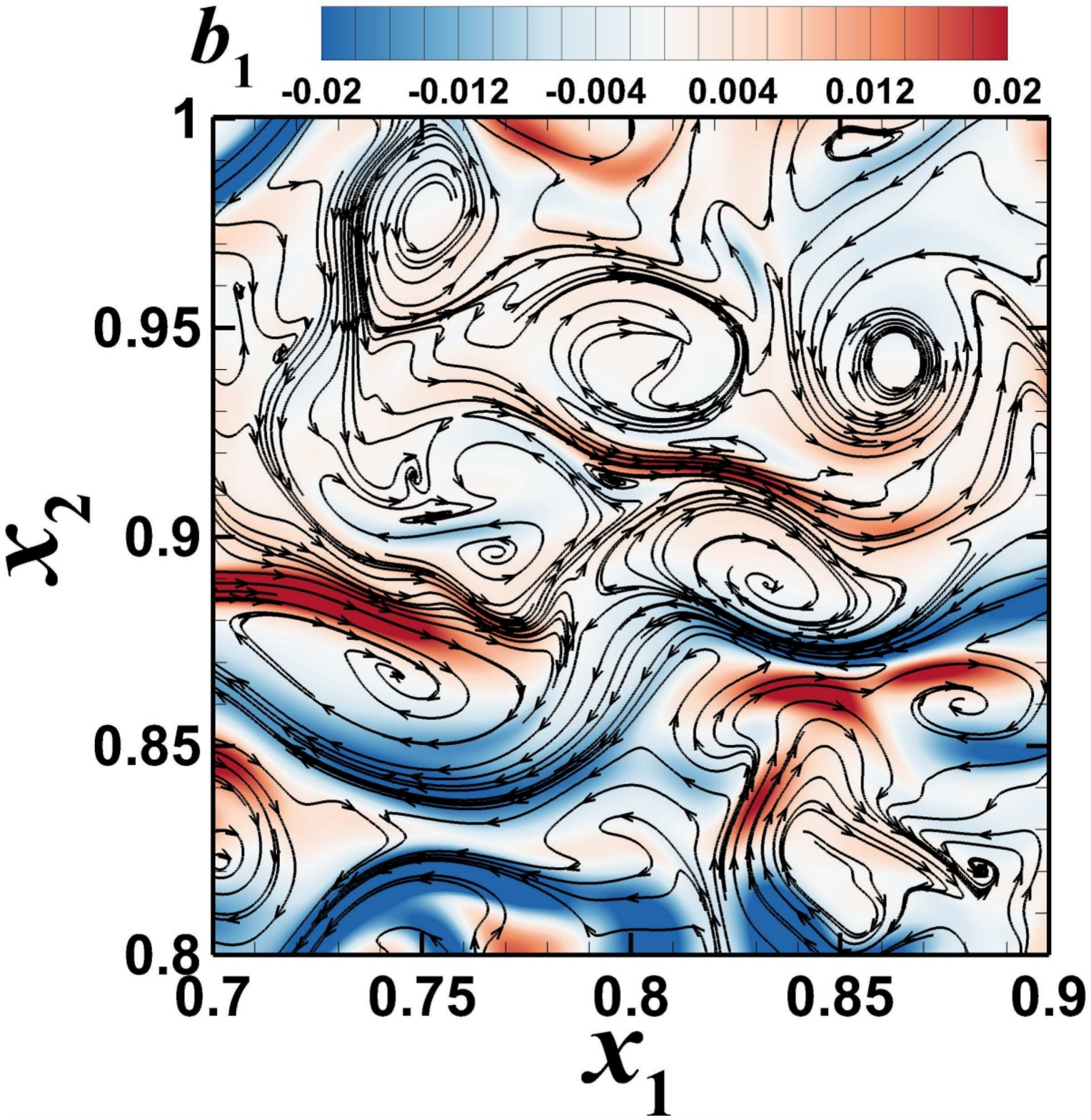}\\
(c) \includegraphics[width=0.4815\linewidth,trim={0.4cm 3.5cm 0.0cm 5cm},clip]{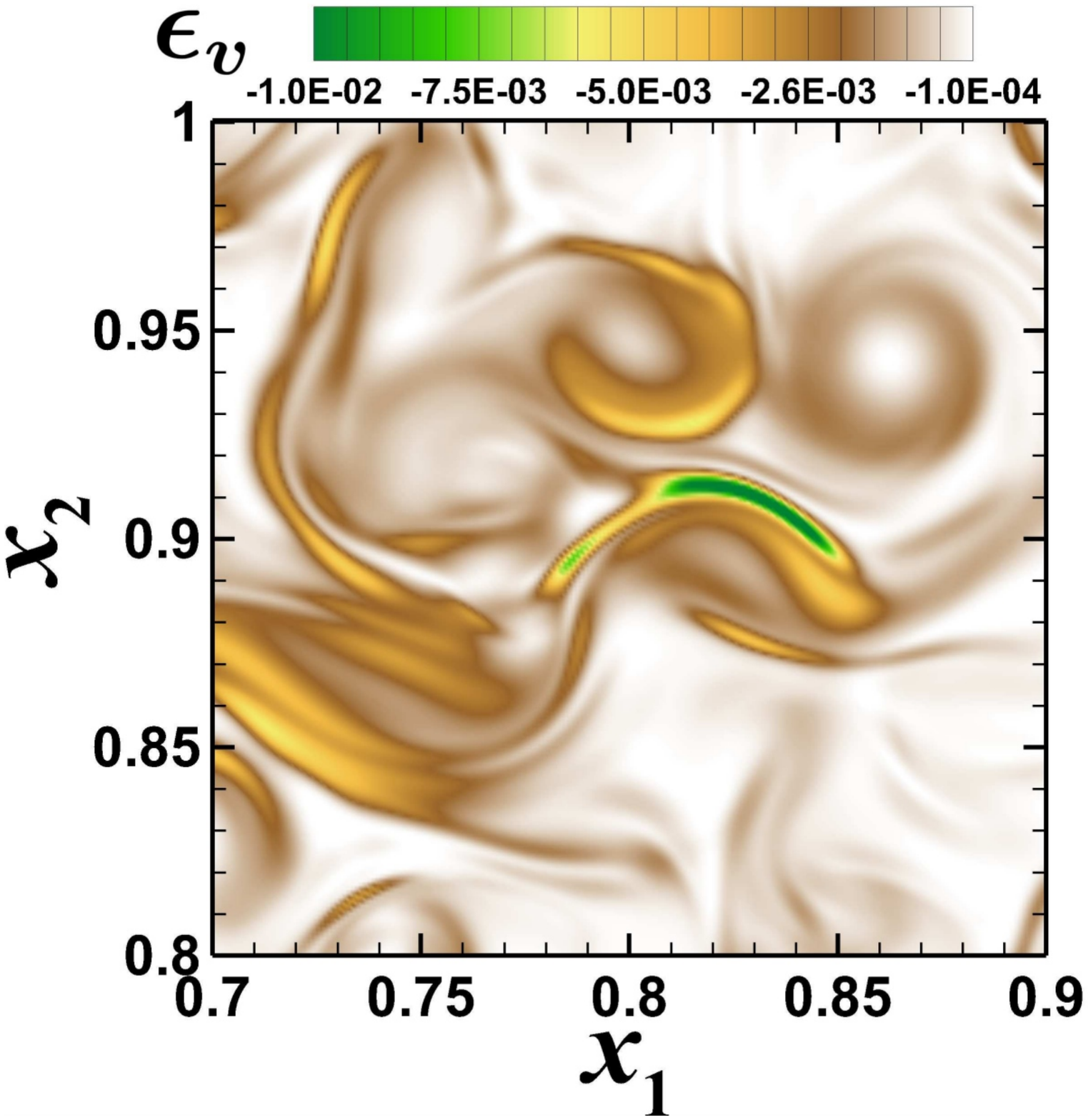}
(d) \includegraphics[width=0.4485\linewidth,trim={2.0cm 3.5cm 0.0cm 5cm},clip]{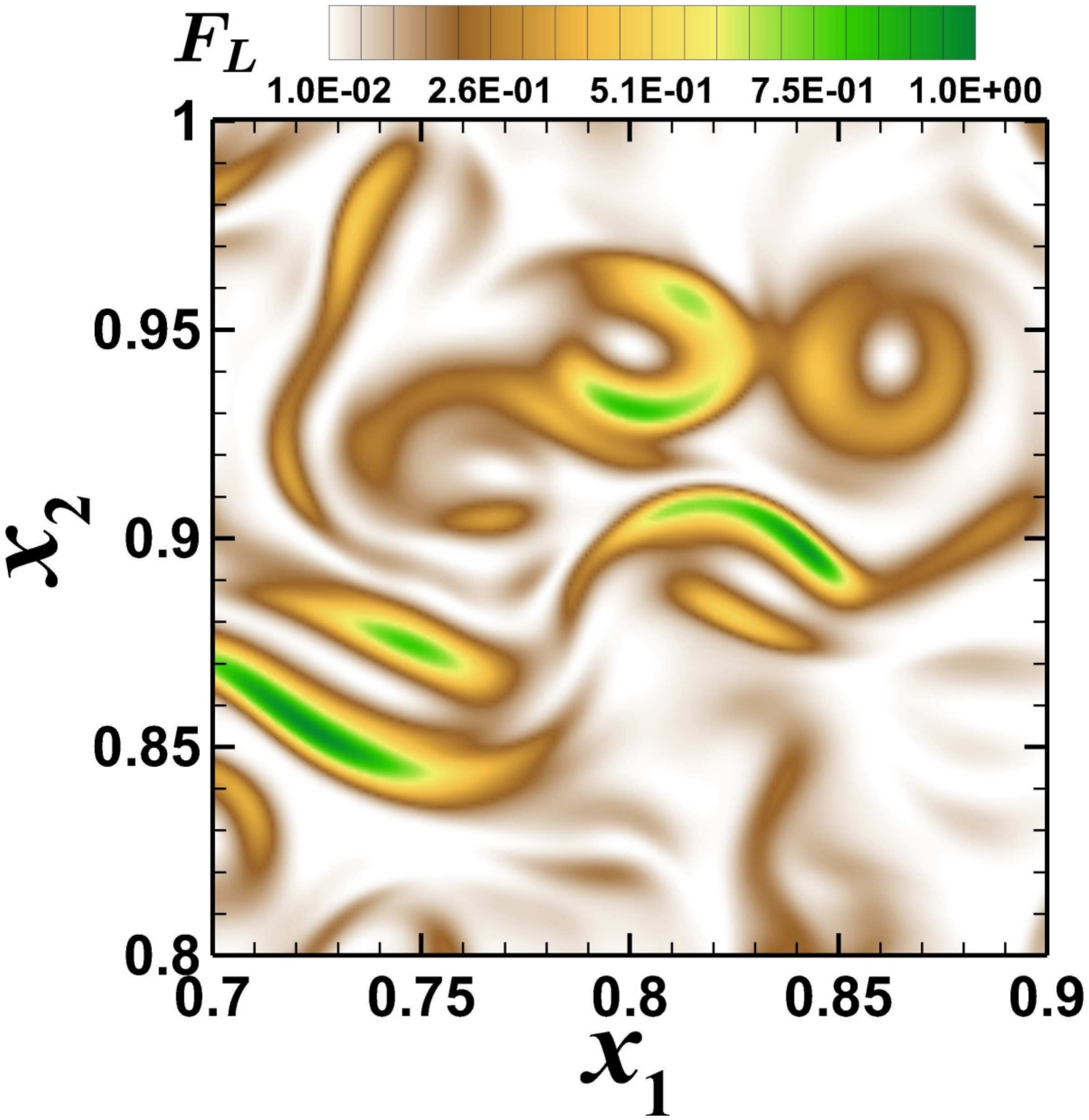}
\caption{Instantaneous snapshots for horizontal planes of (\textit{a}) vertical vorticity, (\textit{b}) magnetic field in $x_{1}$-direction superimposed with magnetic field lines, (\textit{c}) viscous dissipation, and (\textit{d}) Lorentz force magnitude. A quarter of the plane is plotted inside the thermal boundary layer ($x_{3}=-0.495$) for no-slip DC simulation with $\mathcal{R}=3$ in (\textit{a}). The section marked by the black dashed line in (\textit{a}) is enhanced in (\textit{b}), (\textit{c}) and (\textit{d})}
\label{fig:pln}
\end{figure*}

We now look into the reason for the enhancement of Lorentz force at the thermal boundary layer for DC at $\mathcal{R}=3$ with no-slip boundary condition (figure \ref{fig:forces}d). A quarter of the horizontal plane inside the thermal boundary layer $x_{3}=-0.495$ is plotted in figure \ref{fig:pln} to depict the instantaneous vertical vorticity ($\omega_{3}$). The section surrounded by the black dashed line in the vorticity contour plot (figure \ref{fig:pln}a) is focused in figures \ref{fig:pln}b,c and d to show the magnetic field strength in $x_{1}$-direction (i.e., $b_{1}$, superimposed with magnetic field lines), the viscous dissipation and the Lorentz force magnitude respectively. Comparison of the vorticity and the dissipation contours (figure \ref{fig:pln}a and c) exhibit the interaction among vortices that leads to shear layers with intense dissipation around the core of the vortices. The magnetic field lines are observed to concentrate around the edges of these vortices (figure \ref{fig:pln}b). This behavior of the magnetic field can be understood from the flux expulsion mechanism \citep{weiss_1966,gilbert_2016}, where magnetic field lines are advected towards the edges of a vortex. These magnetic field lines around the vortices are stretched due to shear leading to the amplification of the magnetic field, hence resulting in a localized increase (figure \ref{fig:pln}d) in the Lorentz force.\\ 

\begin{figure*}
\centering
(a)\includegraphics[width=0.47\textwidth,trim={0.2cm 6cm 0cm 7.2cm},clip]{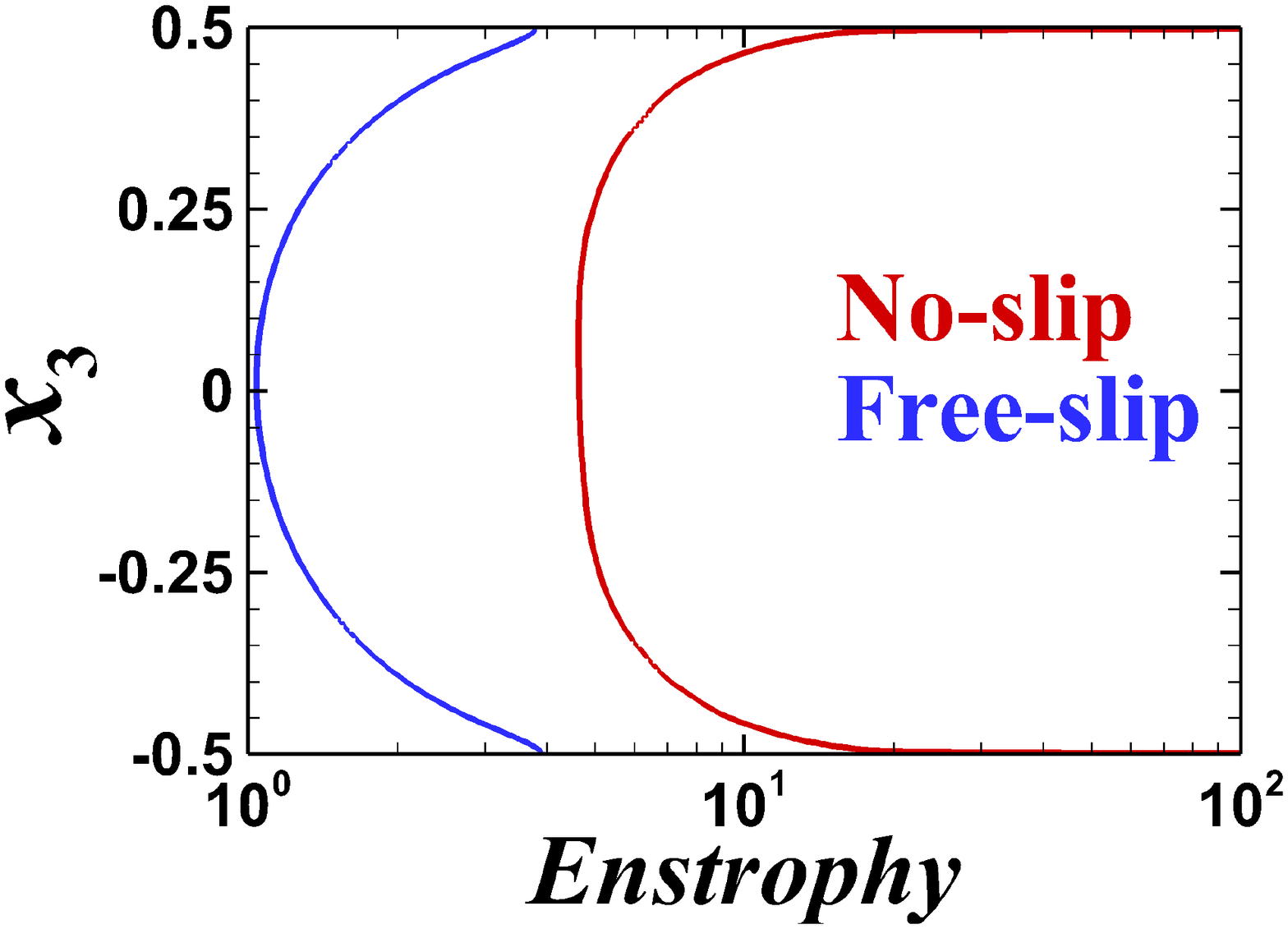}
(b)\includegraphics[width=0.47\textwidth,trim={0.2cm 6cm 0cm 7.2cm},clip]{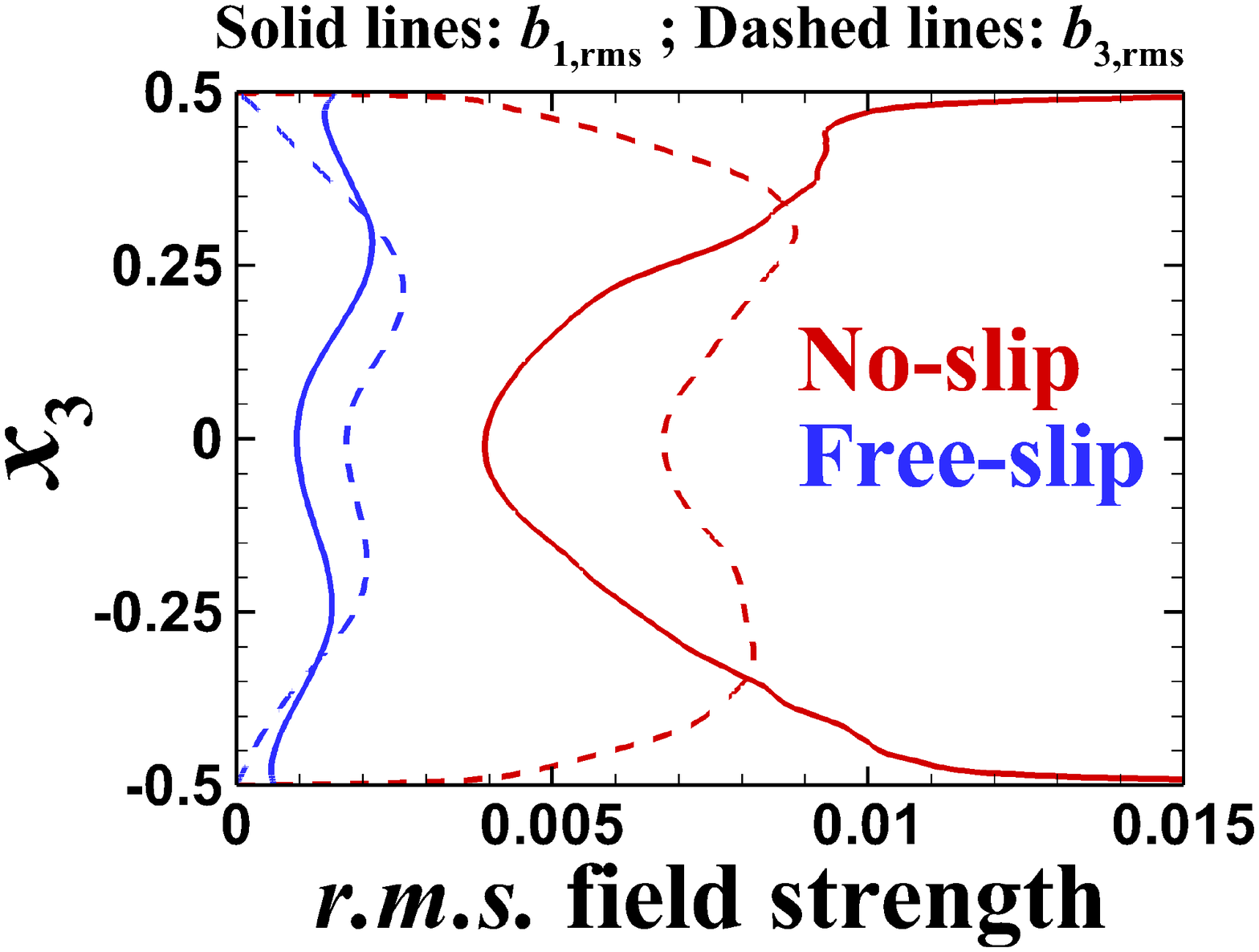}
\caption{Comparison of results for DC at $\mathcal{R}=3$ between no-slip and free-slip conditions: (a) enstrophy and (b) \textit{r.m.s} field strength. Vertical variations are presented after averaging all quantities in the horizontal planes.}
\label{fig:dynamo}
\end{figure*}

Interestingly, the dynamo behavior for DC at $\mathcal{R}=3$ with no-slip and free-slip boundary conditions are found to be very different. In figure \ref{fig:dynamo}a strength of the vortices (represented by the horizontally averaged enstrophy, $1/2\ \overline{\omega_{i}\omega_{i}}$) are an order of magnitude higher near the walls for no-slip boundary condition case as compared to the free-slip boundary condition case. This increase may be attributed to Ekman pumping \citep{hopfinger_1993}. Apart from higher vertical \textit{r.m.s} magnetic field strength ($b_{3,rms}$), an amplification in horizontal magnetic field strength in $x_{1}$ direction ($b_{1,rms}$) is also observed near the wall for the case with no-slip boundary condition in comparison to the case with free-slip boundary condition (see figure \ref{fig:dynamo}b). Because of higher enstrophy, the stretching of magnetic field lines by these vortices, and consequently the increase in the Lorentz force near the walls, are higher with no-slip boundary conditions as compared to free-slip boundary conditions (figure \ref{fig:forces}d,f). This increase in Lorentz force leads to magnetorelaxation in the thermal boundary layer that promotes turbulence and heat transfer enhancement. We do not observe such enhancement in heat transport with free-slip boundary conditions. It should be noted here that for perfectly conducting boundaries, the \textit{r.m.s.} magnetic field has to become parallel to the plates with $b_{3,rms}\xrightarrow{}0$ near the walls (figure \ref{fig:dynamo}b). As the field can not escape the boundaries, a build-up of Lorentz force can be expected \citep{stpierre_1993}. Investigations with insulated wall boundary conditions will be carried out in the future to check whether this mechanism is exclusive to the present combination of boundary conditions.\\ 

\section{Conclusions}\label{sec:conclusion}

We have performed turbulence resolving simulations of the plane layer dynamo. Our study reveals the presence of an optimum thermal forcing at which heat transfer in DC is most efficient compared to RC. The novelty of this investigation is in reporting a local increase in the Lorentz force at the thermal boundary layer edge (figure \ref{fig:forces}d) to be the cause of this enhanced heat transport instead of a global balance between the Lorentz, and the Coriolis forces. This local increase in the Lorentz force subsides the turbulence stabilizing effect of the Coriolis force, and can be viewed as local magnetorelaxation at the thermal boundary layer. This local magnetorelaxation phenomenon results in an increase in the population of plumes, and columnar structures (figure \ref{fig:rhop}c) that increase the intensity of turbulence, and kinetic energy dissipation, especially near the walls (figure \ref{fig:vjd}). We also demonstrate that the stretching of the magnetic field lines around the vortices near the wall is the reason for this enhanced Lorentz force in the thermal boundary layer (figure \ref{fig:pln}). Therefore, the present simulations reveal that capturing the boundary layer dynamics is imperative for understanding the heat transfer characteristics in dynamo convection, and therefore, further investigation on the near-wall behavior of dynamos is warranted.\\


It should be reiterated here that the traditional definition of Elsasser Number, $\Lambda$, is unsuitable for dynamo simulations \citep{soderlund_2015,aurnou_2017,calkins_2018}. The Elsasser Number in a dynamo simulation is not an input parameter that can be externally controlled, unlike the low $Re_m$ magnetoconvection experiments. Therefore, in this study, we choose to plot the heat transfer variation against $\mathcal{R}$, which is more common for heat transfer studies on scaling laws of convection with or without rotation or magnetic field \citep{plumley_2019}. The enhancement of heat transfer at lower values of $\mathcal{R}$ indicates a change of scaling law in strongly rotating convection due to the dynamo action. This parameter choice will help in designing future experiments to facilitate the search for heat transfer scaling laws for dynamo convection.\\


\backsection[Acknowledgements]{The computational support provided by PARAM Sanganak under the National Supercomputing Mission, Government of India at the Indian Institute of Technology, Kanpur is gratefully acknowledged.}


\backsection[Declaration of interests]{ The authors report no conflict of interest.}



\backsection[Author contributions]{  The authors contributed equally to analyzing data and reaching conclusions, and in writing the paper.}

\bibliographystyle{jfm}
\bibliography{jfm}

\end{document}